\documentclass[useAMS,usenatbib,usegraphicx]{mn2e}

\title[An unbiased {H~\sc{i}} study of NGC 3166/9]{Pre-existing dwarfs, tidal knots and a tidal dwarf galaxy: an unbiased {H~\sc{i}} study of the gas-rich interacting galaxy group NGC 3166/9}
\author[K. Lee-Waddell et al.]
	{K.~Lee-Waddell$^{1}$\thanks{E-mail: Karen.Lee-Waddell@rmc.ca (KLW)}, 
	K.~Spekkens$^{1}$, M. P. Haynes$^{2}$, S. Stierwalt$^{3}$, J. Chengalur$^{4}$, 
\newauthor 
	P.~Chandra$^{1}$ and R.~Giovanelli$^{2}$\\
$^{1}$Department of Physics, Royal Military College of Canada, PO Box 17000, Station Forces, Kingston, ON K7K 7B4, Canada\\
$^{2}$Center for Radiophysics and Space Research, Space Sciences Building, Cornell University, Ithaca, NY 14853, USA\\
$^{3}$Spitzer Science Center, California Institute of Technology, 1200 E California Blvd, Pasadena, CA 91125, USA\\
$^{4}$National Centre for Radio Astrophysics, Tata Institute of Fundamental Research, Pune 411 007, India}

\begin{document}

\date{Accepted 2012 September 11. Received 2012 September 10; in original form 2012 April 7}

\pagerange{\pageref{firstpage}--\pageref{lastpage}} \pubyear{2012}

\maketitle

\label{firstpage}

\begin{abstract}
We present Arecibo Legacy Fast ALFA (ALFALFA) and follow-up Giant Metrewave Radio Telescope (GMRT) {H~\sc{i}} observations of the gas-rich interacting group NGC 3166/9.  The sensitive ALFALFA data provide a complete census of {H~\sc{i}}-bearing systems in the group while the high-resolution GMRT data elucidate their origin, enabling one of the first unbiased physical studies of gas-rich dwarf companions and the subsequent identification of second-generation, tidal dwarf galaxies in a nearby group.  The ALFALFA maps reveal an extended {H~\sc{i}} envelope around the NGC 3166/9 group core, which we mosaic at higher resolution using six GMRT pointings spanning $\sim$1 square degree.  A thorough search of the GMRT datacube reveals eight low-mass objects with gas masses ranging from $4 \times 10^7$ to $3 \times 10^8$ M$_{\odot}$ and total dynamical masses up to $1.4 \times 10^9$ M$_{\odot}$.  A comparison of the {H~\sc{i}} fluxes measured from the GMRT data to those measured in the ALFALFA data suggests that a significant fraction ($\sim 60\%$) of the {H~\sc{i}} is smoothly distributed on scales greater than an arcminute ($\sim$7 kpc at the NGC 3166/9 distance).  We compute stellar masses and star formation rates for the eight low-mass GMRT detections, using ancillary SDSS and GALEX data, and use these values to constrain their origin.  Most of the detections are likely to be either pre-existing dwarf irregular galaxies or short-lived, tidally formed knots; however, one candidate, AGC 208457, is clearly associated with a tidal tail extending below NGC 3166, exhibits a dynamical to gas mass ratio close to unity and has a stellar content and star formation rate that are broadly consistent with both simulated as well as candidate tidal dwarf galaxies from the literature.  Our observations therefore strongly suggest that AGC 208457 is a tidal dwarf galaxy.
\end{abstract}



\section{Introduction}
Many present-day galaxies can be found in group environments where tidal interactions within these systems play important roles in galactic dynamics (e.g.~\citealt{t2008}).  Certain interactions in gas-rich groups can form tidal bridges and tails as well as second-generation ``tidal" dwarf galaxies (TDGs), which differ from first-generation ``pre-existing'' dwarfs by their lack of dark matter and higher metallicity content \citep{h2000a}.  Since TDGs form from the material from the outer disks of larger galaxies, they should have total to baryonic mass ratios close to unity  (\citealt{b1992}; \citealt{b2004}).  Not only do TDGs contain populations of recently formed stars that were produced during the interaction event, but they also have older pre-enriched stellar populations \citep{d2000}.  Overall, the formation of any tidal feature or galaxy greatly constrains the type of interaction and the properties of the original objects involved in the process \citep{d2011}.  In groups, the prevalence of tidal galaxies relative to their pre-existing counterparts probes the mechanisms that drive galaxy evolution in these environments \citep{b2006}.

Although TDGs are frequently produced in galaxy interaction simulations and over the last two decades several probable candidates have been identified in $\sim$20 interacting systems, very few of these objects have been widely accepted as authentic (see \citealt {w2003}; \citealt{she2009}).  Tidally formed knots that assemble along tidal filaments tend to have total masses between $10^{6}$ M$_{\odot}$ and $10^8$ M$_{\odot}$; however, in order to survive into long-lived dwarf galaxies -- with life-times $>2$ Gyr -- these objects typically require a total mass $\geq10^8$ M$_{\odot}$ \citep{b2006}.  Models predict that the most probable TDGs are found in the high density tips of tidal tails where large tidal knots are typically formed \citep{b2004}.  Genuine TDGs should also be self-gravitating \citep{d2007}.  Nevertheless, even high-resolution studies of the neutral hydrogen ({H~\sc{i}}) dynamics of well-known TDG candidates in the southern tail of NGC 4038/9 failed to prove that the objects are rotating -- a clear indicator of self-gravitation -- and that they are therefore kinematically distinct from the tidal tail, which has given rise to much debate and speculation about their nature \citep{h2001}.  As can be seen, deciding whether a given object is a tidal dwarf or not is a non trivial issue and generally requires a host of corroborating observations.

Given the properties of (young) TDGs as well as the environments within which they form, {H~\sc{i}} observations are a reasonable preliminary search tool.  {H~\sc{i}} traces the location of tidally formed features, can indicate regions of potential star formation and has been routinely used to map the gas distribution in and around gas-rich systems (see \citealt{f2009a}; \citealt{k2009}; \citealt{s2009}; \citealt{b2010a} for recent examples).  {H~\sc{i}} measurements can be used to estimate gas masses and total masses (discussed further in $\S$ 4).  With sufficient resolution, the internal structure and dynamics of TDG candidates can also be constrained \citep{d2007}.  Interferometric data from instruments such as the Very Large Array (VLA; e.g.~\citealt{f2009}) and the Australia Telescope Compact Array (ATCA; e.g.~\citealt{p2011}) become essential for resolving the structure of more compact features.  The Faint Irregular Galaxies GMRT Survey (FIGGS; \citealt{b2008}) shows that the Giant Metrewave Radio Telescope (GMRT) is also particularly useful in surveying gas-rich dwarfs as its fixed antenna configuration allows for maps at different angular resolutions to be produced by tapering the data during imaging.

Considerable effort has been devoted to interferometric {H~\sc{i}} mapping of nearby interacting systems in order to search for TDGs (e.g.~\citealt{d2000}; \citealt{h2001}; \citealt{b2007}).  Nevertheless, none are blind in their approach: only the regions most likely to harbour second-generation objects were studied in detail, thus probing a incomplete view of the dwarf galaxy population in the groups.  By contrast, single-dish surveys such as the {H~\sc{i}} Parkes All-Sky Survey (HIPASS; \citealt{k2004}), the Arecibo Legacy Fast ALFA survey (ALFALFA; \citealt{g2005}) and those using the Green Bank Telescope (GBT; e.g.~\citealt{b2010}) can efficiently cover the large sky areas needed to characterize group environments but lack the angular resolution to probe the structure of individual objects. An unbiased census of TDGs in nearby groups is therefore enabled by a two-pronged approach: sensitive single-dish observations to locate {H~\sc{i}}-bearing systems on degree scales and high-resolution interferometric follow-up to elucidate their origin. In this paper, we target low-mass ALFALFA {H~\sc{i}} detections in the nearby gas-rich group NGC 3166/9 with the GMRT affording one of the first blind, systematic studies of both first- and second- generation systems in a group environment.  

NGC 3166/9 consists of three main spiral galaxies: NGC 3165, NGC 3166 and NGC 3169 and is located in the Sextans constellation.  The heliocentric velocity difference between these galaxies is small -- $cz_{\odot}$ =  1340 $\pm$ 5 km s$^{-1}$ \citep{s1990}, 1345 $\pm$ 5 km s$^{-1}$ \citep{v2001} and 1238 $\pm$ 4 km s$^{-1}$ \citep{v2001} for NGC 3165, NGC 3166 and NGC 3169 respectively -- and they share a common {H~\sc{i}} distribution at arcminute resolution (see below).  The close proximity of these galaxies to one another as well as a tail-like feature extending below NGC 3166 in the ultraviolet (UV; discussed in $\S$ 5), indicates previous and on-going interactions within the group.  The average distance to NGC 3169, derived from the type 1A supernova SN2003cg, is 22.6 Mpc (\citealt{w2008}; \citealt{m2009}), which is assumed for all group members throughout this paper.

Recently, the extended {H~\sc{i}} emission of this group, first observed by \citet{h1981}, was mapped at 4$\arcmin$ (30 kpc at the distance of NGC 3169) resolution by ALFALFA (\citealt{g2005}; Fig.~\ref{ALFA}).  We present the ALFALFA data for the NCG 3166/9 region in $\S$ 2.  Most of the detected {H~\sc{i}} is located around NGC 3169 with various features extending outwards; however, the spatial resolution of the ALFALFA data is too low to characterize the structure of individual group members.  The high {H~\sc{i}} content and clear signs of interactions in the NGC 3166/9 group make it an ideal candidate for high-resolution follow-up to distinguish first- and second-generation objects, which should result in a better understanding of galaxy evolution in groups.

We obtained interferometric {H~\sc{i}} observations using the GMRT, as a follow-up to the ALFALFA data, in an attempt to resolve the putative dwarf galaxy population in NGC 3166/9.  We have been able to make higher resolution maps of the {H~\sc{i}} features resulting in the detection of several low-mass objects ($\S$ 4).  We computed {H~\sc{i}} masses and total masses, from the GMRT data, for each detection ($\S$ 4).  Stellar masses and star formation rates (SFRs) were also estimated for each object using ancillary optical and UV observations ($\S$ 5).  The combination of mass ratio estimates, SFRs and other galactic properties have been used to categorize the low-mass {H~\sc{i}} features and identify a probable TDG in our neutral/blind survey of the gas-rich objects in NGC 3166/9 ($\S$ 6).


\section[]{ALFALFA Observations of NGC 3166/9}

ALFALFA mapping of the NGC 3166/9 group was completed during spring 2009.  The ALFALFA data acquisition and reduction strategy are described in detail in \citet{h2011} and references therein.  Processing of the NGC 3166/9 data was carried out as described in \citet{h2011} but the resulting {H~\sc{i}} detections were not included in the ALFALFA 40\% catalogue ($\alpha$.40) presented in that paper because NGC 3166/9 lies outside the $\alpha$.40 sky footprint.  We therefore present the ALFALFA detections in the NGC 3166/9 region here.

\begin{figure}
  \includegraphics[width=84mm]{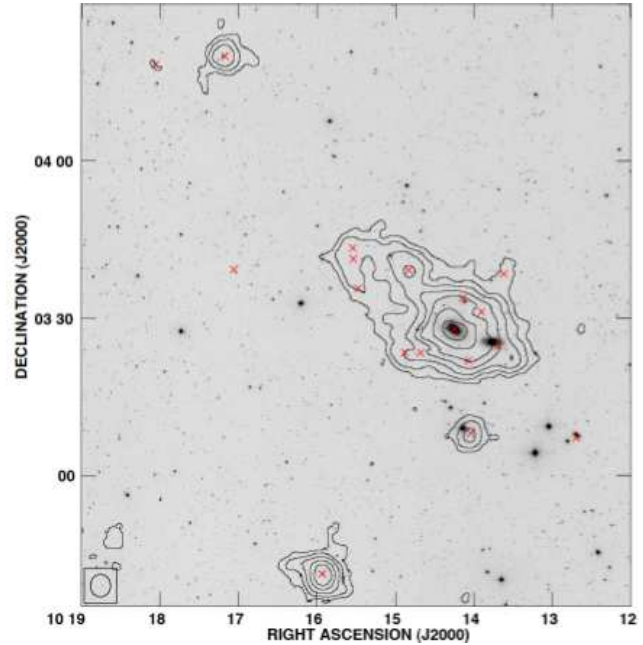}
  \caption{ALFALFA total {H~\sc{i}} intensity contours superimposed on a SDSS \textit{r}-band greyscale image of the region surrounding the NGC 3166/9 group.  The 4$\arcmin$ ALFALFA beam is in the bottom left corner and the ``X''s indicate the locations of the ALFALFA {H~\sc{i}} detections that have been identified using an automated source extractor.  The intensity contours are at $N_{H_I} = (1.125, 2.25, 4.5, 9, 15, 30, 60) \times 10^{19}$ atoms cm$^{-2}$.  Due to the varying noise in the cube, the intensities associated with AGC 5503 and AGC 204302 (see Table~\ref{ALF_detect}) fall below the lowest contour in the moment map. \label{ALFA}}
\end{figure}

A total intensity {H~\sc{i}} map of the emission detected by ALFALFA in the NGC 3166/9 region is shown in Fig.~\ref{ALFA}.  ALFALFA detects an extended {H~\sc{i}} envelope around NGC 3165, NGC 3166 and NGC 3169 with a maximum diameter of 40$\arcmin$ (300 kpc) at column densities N$_{H_I}$ $\sim 1 \times 10^{19}$ cm$^{-2}$, as well as several sources in the periphery with $989 < cz_{\odot} < 1484$ km s$^{-1}$.  The ALFALFA source extractor \citep{s2007} finds a total of eighteen statistically significant emission line sources, some of which may be sub-condensations of a larger structure, in the region $10\textsuperscript{h}12\textsuperscript{m} < \alpha < 10\textsuperscript{h}19\textsuperscript{m}$, $+02^{\circ}30\arcmin < \delta < +04^{\circ}30\arcmin$ and their properties are presented in Table~\ref{ALF_detect}.  The definitions of these measured parameters are identical to those in table 1 of \citet{h2011} and brief descriptions are reproduced in the notes to Table~\ref{ALF_detect} for clarity.  To distinguish them from parameters measured in our interferometric maps, all parameters derived from ALFALFA data are primed.  Each ALFALFA detection has been assigned an Arecibo General Catalogue (AGC) number.  Only four of the detections in this region are associated with previously catalogued optical counterparts.  Conversely, there are three faint optical dwarf galaxy candidates (not included in Table~\ref{ALF_detect}) in the SDSS image that fall within the {H~\sc{i}} envelope in Fig.~\ref{ALFA} but for which no localized {H~\sc{i}} emission is detected.  The fourteen new AGC objects in Table~\ref{ALF_detect} have $M_{H_I} < 10^9$ M$_{\odot}$ at the NGC 3166/9 distance and many lie within the group core.  Of particular interest is AGC 208457, a relatively bright {H~\sc{i}} source with a faint optical counterpart located at the tip of tail-like structures discernible in the optical and in the UV (see $\S$ 5); its location in the group suggests a tidal origin.  To further understand the nature of this source and the other low-mass ALFALFA detections, we obtained follow-up GMRT observations of the NGC 3166/9 group core.

\begin{table*}
\centering
 \begin{minipage}{170mm}
\caption{ALFALFA Detections}
\label{ALF_detect}
\begin{tabular}{@{}l c c c r@{}l r@{}l r@{}l c c} 
\hline
AGC	&Name		&{H~\sc{i}} Coordinates	&Optical Coordinates		&\multicolumn{2}{c}{$cz'_{\odot}$}		&\multicolumn{2}{c}{W$'_{50}$}	&\multicolumn{2}{c}{S$'_{21}$}			&S/N$'$
&$\sigma'$\\
		&			&(J2000)				&(J2000)				&\multicolumn{2}{c}{(km s$^{-1}$)}	&\multicolumn{2}{c}{(km s$^{-1}$)}	&\multicolumn{2}{c}{(Jy km s$^{-1}$)}	&	
&(mJy)	\\
(1)		&(2)			&(3)						&(4)						&\multicolumn{2}{c}{(5)}			&\multicolumn{2}{c}{(6)}			&\multicolumn{2}{c}{(7)}				&(8)	
&(9)		\\
\hline
5503	&NGC 3156	&10 12 42.0,	+03 07 10	&10 12 41.3,	+03 07 45	&1279	&\hspace{1mm}$\pm$ 29	&200	&\hspace{1mm}$\pm$ 57	&0.98 	&\hspace{1mm}$\pm$ 0.13	&4.3
&3.58	\\
208533	&			&10 13 37.1,	+03 38 29	&						&1274	&\hspace{1mm}$\pm$ 1	&27		&\hspace{1mm}$\pm$ 1	&0.99	&\hspace{1mm}$\pm$ 0.03	&14.1
&2.69	\\
5516	&NGC 3166	&10 13 41.4,	+03 24 44	&10 13 45.6,	+03 25 28	&1326	&\hspace{1mm}$\pm$ 4	&193	&\hspace{1mm}$\pm$ 7	&9.85	&\hspace{1mm}$\pm$ 0.12	&54.6
&2.89	\\
208534	&			&10 13 54.3,	+03 31 15	&						&1302	&\hspace{1mm}$\pm$ 1	&51		&\hspace{1mm}$\pm$ 2	&5.37	&\hspace{1mm}$\pm$ 0.05	&58.4
&2.74	\\
208443	&			&10 14 02.4,	+03 08 04	&10 14 01.0,	+03 08 27	&1484	&\hspace{1mm}$\pm$ 1	&50		&\hspace{1mm}$\pm$ 2	&3.22	&\hspace{1mm}$\pm$ 0.06	&40.2
&2.50	\\
208457	&			&10 14 03.9,	+03 21 53	&10 14 08.0,	+03 20 41	&1346	&\hspace{1mm}$\pm$ 1	&44		&\hspace{1mm}$\pm$ 1	&5.78	&\hspace{1mm}$\pm$ 0.04	&76.3
&2.50	\\
208535	&			&10 14 08.3,	+03 33 50	&						&1109	&\hspace{1mm}$\pm$ 1	&41		&\hspace{1mm}$\pm$ 1	&4.48	&\hspace{1mm}$\pm$ 0.04	&62.0
&2.41	\\
5525	&NGC 3169	&10 14 14.7,	+03 27 56	&10 14 14.8,	+03 27 59	&1232	&\hspace{1mm}$\pm$ 2	&452	&\hspace{1mm}$\pm$ 4	&81.5	&\hspace{1mm}$\pm$ 0.18	&244.5
&3.29	\\
208536	&			&10 14 40.7,	+03 23 25	&						&1226	&\hspace{1mm}$\pm$ 1	&33		&\hspace{1mm}$\pm$ 1	&1.76	&\hspace{1mm}$\pm$ 0.04	&24.7
&2.71	\\
204288	&			&10 14 49.8,	+03 39 11	&10 14 51.5,	+03 38 53	&989	&\hspace{1mm}$\pm$ 1	&43		&\hspace{1mm}$\pm$ 3	&7.56	&\hspace{1mm}$\pm$ 0.07	&95.0	
&2.67	\\
208537	&			&10 14 52.8,	+03 23 27	&						&1181	&\hspace{1mm}$\pm$ 1	&30		&\hspace{1mm}$\pm$ 1	&2.34	&\hspace{1mm}$\pm$ 0.03	&35.6
&2.48	\\
208329	&			&10 15 28.4,	+03 35 44	&10 15 31.9,	+03 35 08	&1018	&\hspace{1mm}$\pm$ 2	&47		&\hspace{1mm}$\pm$ 3	&1.15	&\hspace{1mm}$\pm$ 0.05	&15.6
&2.36	\\
208538	&			&10 15 32.1,	+03 41 18	&						&989	&\hspace{1mm}$\pm$ 1	&55		&\hspace{1mm}$\pm$ 3	&3.8	&\hspace{1mm}$\pm$ 0.08	&35.5
&3.19	\\
208539	&			&10 15 32.6,	+03 43 25	&						&975	&\hspace{1mm}$\pm$ 1	&27		&\hspace{1mm}$\pm$ 3	&2.67	&\hspace{1mm}$\pm$ 0.06	&42.3
&2.62	\\
5539	&036-075	&10 15 56.1,	+02 41 10	&10 15 55.1,	+02 41 09	&1278	&\hspace{1mm}$\pm$ 1	&141	&\hspace{1mm}$\pm$ 2	&14.73	&\hspace{1mm}$\pm$ 0.08	&113.8
&2.43	\\
204302	&			&10 17 03.5,	+03 39 17	&10 17 02.3,	+03 38 46	&1039	&\hspace{1mm}$\pm$ 6	&33		&\hspace{1mm}$\pm$ 12	&0.36	&\hspace{1mm}$\pm$ 0.04	&5.9
&2.31	\\
5551	&			&10 17 10.6,	+04 20 01	&10 17 11.8,	+04 19 50	&1344	&\hspace{1mm}$\pm$ 1	&56		&\hspace{1mm}$\pm$ 3	&5.09	&\hspace{1mm}$\pm$ 0.07	&57.8
&2.59	\\
208392	&			&10 18 03.6,	+04 18 24	&10 18 03.7,	+04 18 38	&1322	&\hspace{1mm}$\pm$ 3	&34		&\hspace{1mm}$\pm$ 6	&0.55	&\hspace{1mm}$\pm$ 0.05	&8.8
&2.31	\\
\hline 
\multicolumn{12}{l}{\footnotesize Parameter definitions are identical to those described in detail by \citet{h2011}.  For clarity, all parameters derived from}\\
\multicolumn{12}{l}{\footnotesize the ALFALFA data are primed.}\\
\multicolumn{12}{l}{\footnotesize Col. (1) entry number in the Arecibo General Catalog (AGC), a private database of extragalactic objects maintained by }\\
\multicolumn{12}{l}{\footnotesize \hspace{1cm} M.P. Haynes and R. Giovanelli}\\
\multicolumn{12}{l}{\footnotesize Col. (2) common name of the associated optical counterpart, where applicable}\\
\multicolumn{12}{l}{\footnotesize Col. (3) centroid of the {H~\sc{i}} line source}\\
\multicolumn{12}{l}{\footnotesize Col. (4) centroid of the most probable optical counterpart, where applicable}\\
\multicolumn{12}{l}{\footnotesize Col. (5) heliocentric velocity of the {H~\sc{i}} source}\\
\multicolumn{12}{l}{\footnotesize Col. (6) velocity width of the line profile measured at the 50\% level of the peak flux}\\
\multicolumn{12}{l}{\footnotesize Col. (7) integrated {H~\sc{i}} line flux density}\\
\multicolumn{12}{l}{\footnotesize Col. (8) signal-to-noise}\\
\multicolumn{12}{l}{\footnotesize Col. (9) RMS noise of the spatially integrated spectral profile}\\
\end{tabular}
\end{minipage} 
\end{table*}


\section[]{GMRT Observations and Data Reduction}
\begin{figure}
  \includegraphics[width=84mm]{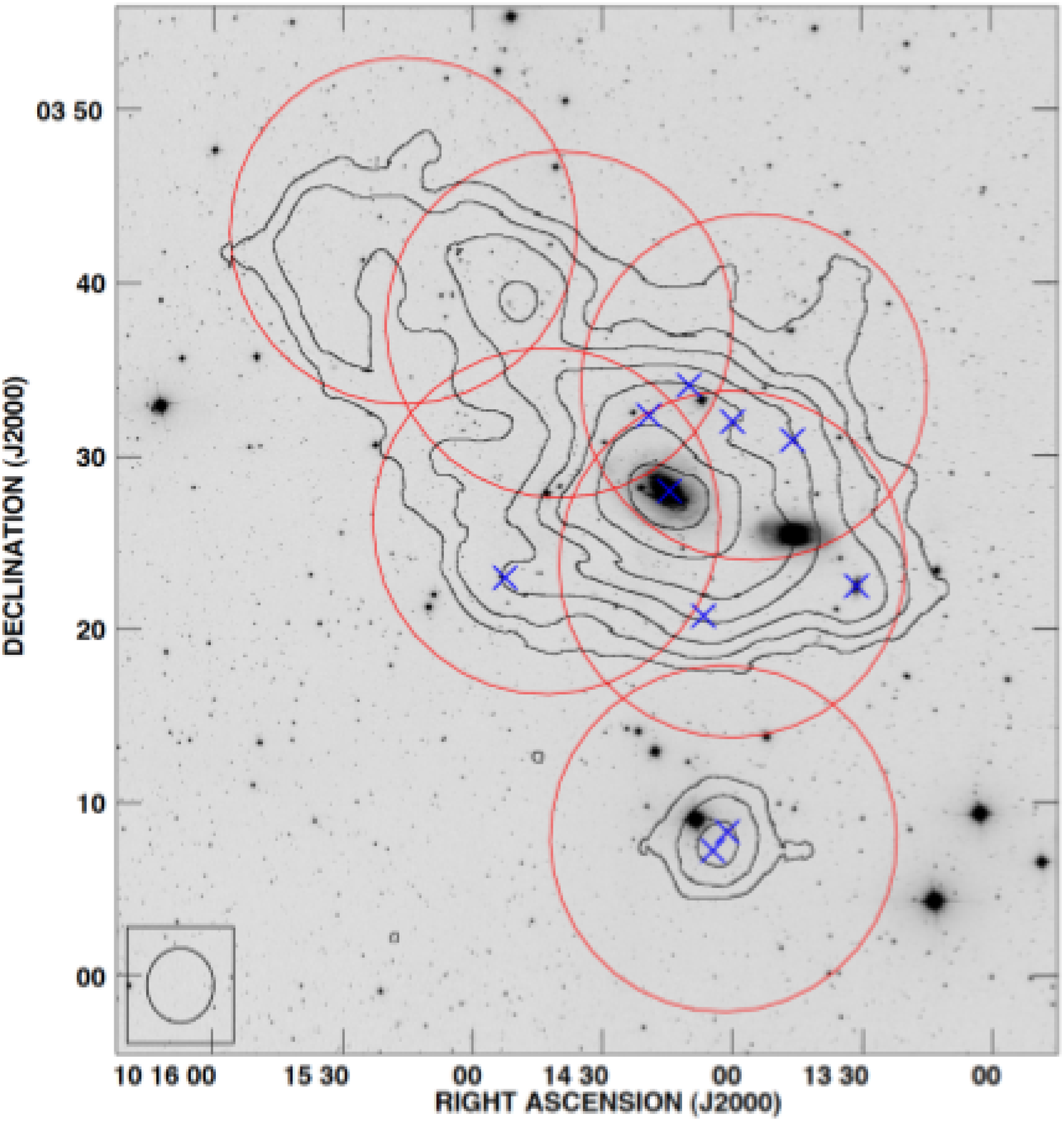}
  \caption{GMRT follow-up pointings (red circles) superimposed on the ALFALFA total intensity contours of the core region of the NGC 3166/9 group.  The contours shown are the same as in Fig.~\ref{ALFA}.  The ``X''s indicate the locations of the GMRT {H~\sc{i}} detections that will be discussed in $\S$ 4.  \label{point}}
\end{figure}

The GMRT observations consisted of six pointings -- each with a half-power beam width (HPBW) of $\sim$20$\arcmin$ -- with five pointings of the core region of NGC 3166/9 and one pointing of a nearby {H~\sc{i}} source located at $\alpha =$ 10\textsuperscript{h}14\textsuperscript{m}00\textsuperscript{s}, $\delta = +03^{\circ}08\arcmin00\arcsec$, observed over a total of five nights (Fig.~\ref{point}).  The 40 hours of telescope time, which included calibration time, were divided between the pointings to ensure that there was sufficient sensitivity for all regions with a focus on the ALFALFA detections in Table~\ref{ALFA}.  At the time of the observations, the GMRT software correlator was a fairly new addition to the telescope and provided twice the number of spectral line channels (256 channels total) across a given band for improved editing over the original hardware correlator \citep{r2010}.  Data were collected simultaneously from both correlators.  Once it was determined that the software correlator was operating properly and was producing acceptable results, the data from this correlator were used exclusively.  Standard flux calibrators, 3C147 and 3C286, were observed at the beginning and end of every observing night for 15 minutes each; whereas, a nearby phase calibrator, 0943-083, was observed for 8 minutes for every 45 minutes on the target source.  The observing map parameters are summarized in Table~\ref{param}.

The observations were reduced using the Astronomical Image Processing System (AIPS) version 31Dec10 \citep{g2003}.  A pseudo-continuum map was used for the initial calibration and editing for each night of observations.  After detailed editing of each baseline, the corrections were then applied to the spectral line cube and the visibilities in individual channels were closely examined.  The full 16.3 kHz (3.46 km s$^{-1}$) spectral resolution of the software correlator was retained during editing to
facilitate radio frequency interference (RFI) excision.  In cases where substantial RFI was removed in one channel, the same editing was applied to all channels to ensure a consistent synthesized beam shape across the whole cube.  For all five nights of observations the channel corresponding to 1161 km s$^{-1}$, which for the most part fell in between detections (see $\S$ 4), had an appreciable amount of RFI and was removed entirely.  The data were then calibrated using standard AIPS  tasks where the flux density for each source was determined relative to that of the flux calibrators and the phase of the data was retrieved from the computed phase closures from the phase calibrator.  Bandpass solutions were derived using the flux calibrators.  This data editing and calibration process was applied to each of the six pointings to ensure consistency across the final mosaic.

\begin{table}
 \centering
 \begin{minipage}{90mm}
 \caption{GMRT Observation Setup and Image Properties}
 \label{param}
\begin{tabular}{@{} l c c}  
\hline
Parameter 								& Value		& Units		\\ 
\hline 
Number of pointings						& 6			&			\\
Average time on source per pointing		& 290		& min		\\ 
Primary beam HPBW per pointing			& 19.7		& arcmin		\\
Mosaicked map size						& 80		& arcmin		\\ 
Central observing frequency				& 1414.6	& MHz		\\
Observing bandwidth						& 4			& MHz		\\ 
Final spectral resolution					& 32.6		& kHz		\\ 
Final spectral resolution					& 6.9		& km s$^{-1}$\\ 
Map spatial resolutions					& 15 \& 45	& arcsec		\\ 
Peak map sensitivity (45$\arcsec$ res.)		& 1.64		& mJy beam$^{-1}$\\  
Peak map sensitivity (15$\arcsec$ res.)		& 0.958		& mJy beam$^{-1}$\\  
\hline
\end{tabular} 
\end{minipage}
\end{table}

\begin{figure}
  \includegraphics[width=84mm]{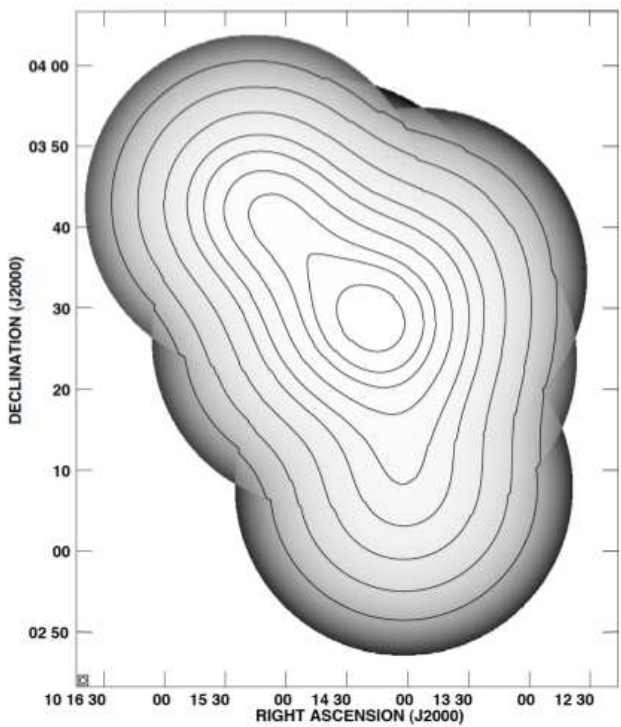}
  \caption{Noise map of the 45$\arcsec$ resolution GMRT mosaic.  The greyscale ranges linearly from 1.64 to 29.4 mJy beam$^{-1}$ and the contours are at $\sigma$ = (1.7, 1.8, 1.9, 2.1, 2.4, 2.8, 3.7, 5.7, 10) mJy beam$^{-1}$. \label{noise}}
\end{figure}

The calibrated datacubes were imaged at a variety of resolutions to emphasize structures on different angular scales.  Images were synthesized using a two-channel spectral average, which included the blanked channel corresponding to 1161 km s$^{-1}$, in order to improve the sensitivity of each channel and to produce a final spectral resolution of 6.9 km s$^{-1}$ (see Table~\ref{param}).  We choose to use datacubes at two complementary angular resolutions in our analysis.  For the datacubes that were used to identify gas-rich objects and measure flux densities, natural weighting of the data produced a $\sim$45$\arcsec$ (5 kpc) synthesized beam and the highest sensitivity.  The cubes that were used to determine the size and resolve the structure of the detected low-mass objects were imaged with a taper that produced a $\sim$15$\arcsec$ (2 kpc) resolution.  Once the datacubes for each pointing were imaged and deconvolved, the pointings were primary beam corrected and interpolated to form a mosaic of the entire region.  The overlapping pointings resulted in a non-uniform noise distribution across the maps as shown in Fig.~\ref{noise} for the 45$\arcsec$ resolution datacube, which has a peak sensitivity of $\sigma$ = 1.64 mJy beam$^{-1}$.  

\section[]{{H~\sc{i}} in NGC 3166/9}

\begin{figure*}
\begin{center}
  \includegraphics[width=84mm]{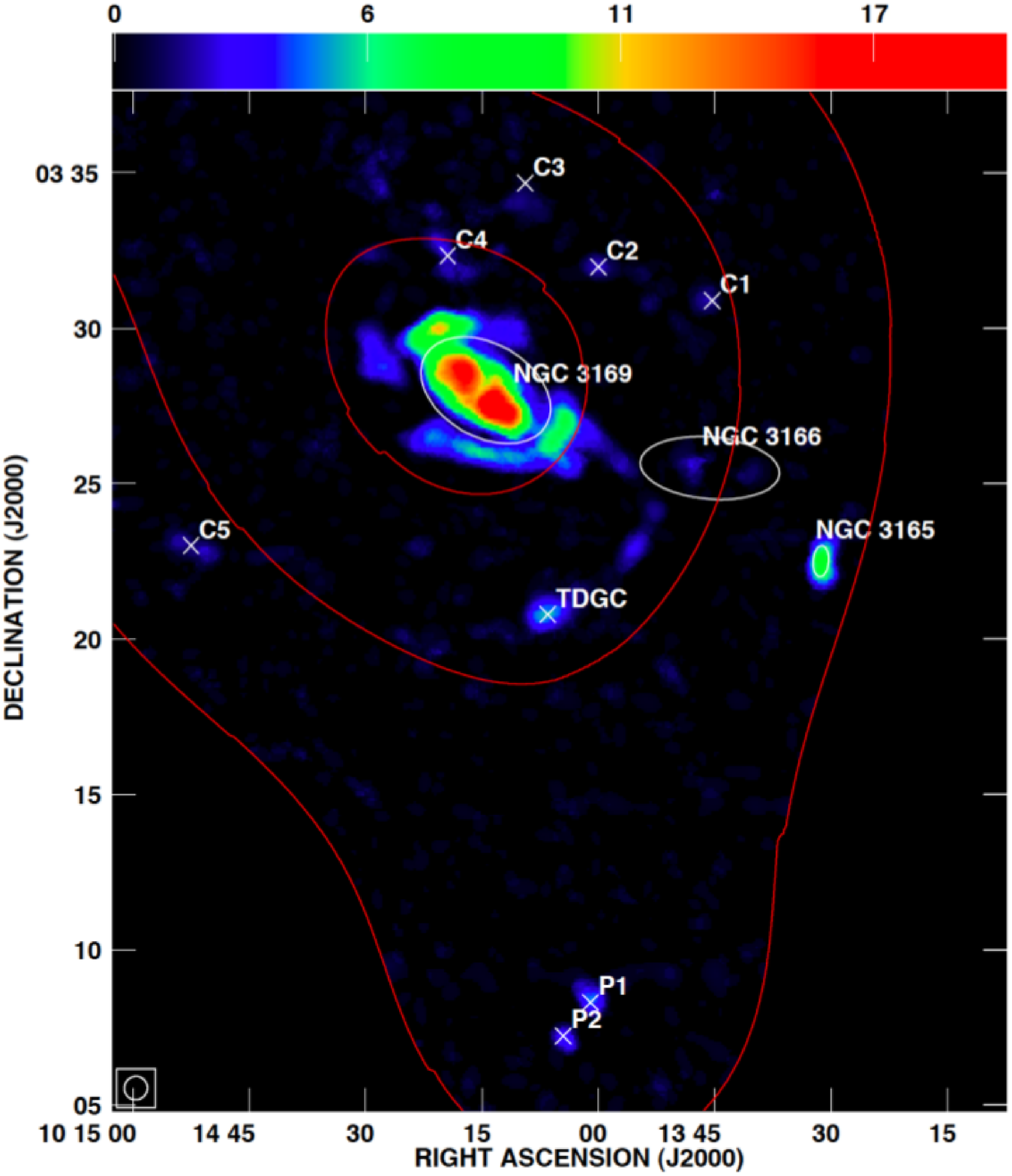}
    \includegraphics[width=84mm]{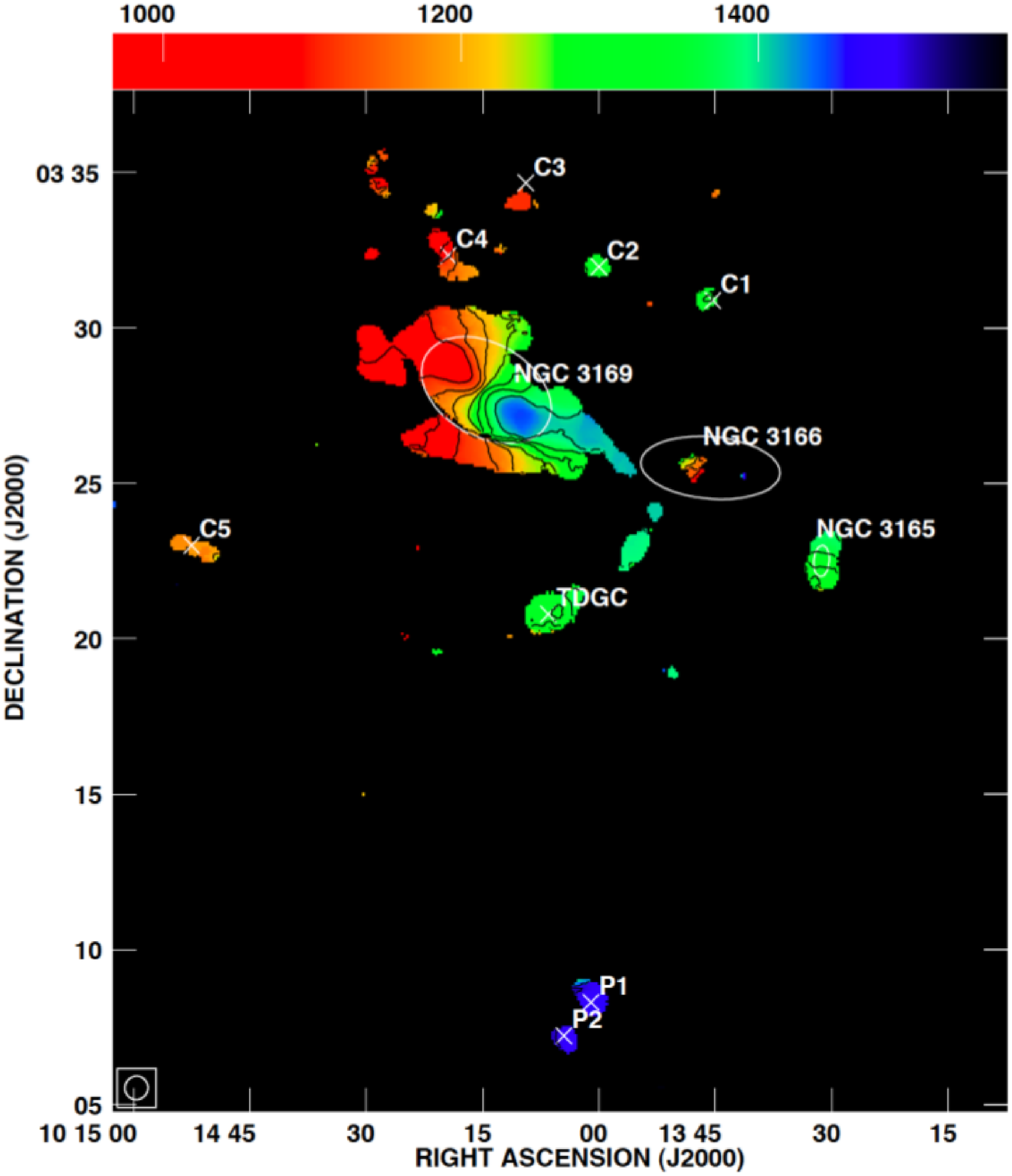}
  \caption{\textit{Left:} GMRT total intensity map of the {H~\sc{i}} in NGC 3166/9 at 45$\arcsec$ resolution.  The red contours show the regions in the map with $\sigma$ = (1.7, 2, 2.8) mJy beam$^{-1}$ and the colour scale ranges linearly from 3 to 19 $\times10^{20}$ atoms cm$^{-2}$.  \textit{Right:} GMRT intensity-weighted velocity map of the {H~\sc{i}} in NGC 3166/9 at 45$\arcsec$ resolution.  Velocity contours are at (1000, 1050, 1100, 1150, 1200, 1250, 1300, 1350, 1400, 1450, 1500) km s$^{-1}$.  In both maps, the synthesized beam is shown in the bottom left corner and the locations of the {H~\sc{i}} detections and the main group of galaxies are labelled.  Ellipses on the NGC galaxies represent the sizes of their optical counterparts.  To remove spurious signals, regions in each datacube channel where the map noise $\sigma > 2.8$ mJy beam$^{-1}$ or where the signal-to-noise S/N $<3$ were blanked before the moments were computed.  In addition, the velocity map is blanked at locations where $N_{H_I} \leq 1.1\times10^{20}$ atoms cm$^{-2}$. \label{HI}}
\end{center}
\end{figure*}

The 45$\arcsec$ mosaicked GMRT datacube was meticulously searched to identify all {H~\sc{i}} detections within the system.  {H~\sc{i}} features considered to be detections were required to appear in at least three consecutive velocity channels with peak flux densities greater than 4.5 times the RMS noise at their location in the cube.  Sources C1 and C4 (defined in Fig.~\ref{HI} and Table~\ref{mass}) are at $\sim$4.7$\sigma$; whereas, all other objects have peak fluxes that exceed $6\sigma$.  Near the centre of the 45$\arcsec$ resolution map, where $\sigma \approx 1.6$ mJy beam$^{-1}$, a source with an {H~\sc{i}} mass $M_{H_I} \geq 2.1 \times 10^7$ $M_{\odot}$ -- corresponding to an {H~\sc{i}} column density $N_{H_I} \geq 9.5 \times10^{19}$ atoms cm$^{-2}$ -- was therefore considered a detection.  Slight variations in these criteria for identifying detections, such as raising the flux density threshold (by 5\%) or increasing the number of channels in which the emission must appear (by 1 channel), do not change the number of reported detections.

\begin{figure*}
\begin{center}
    \includegraphics[width=168mm]{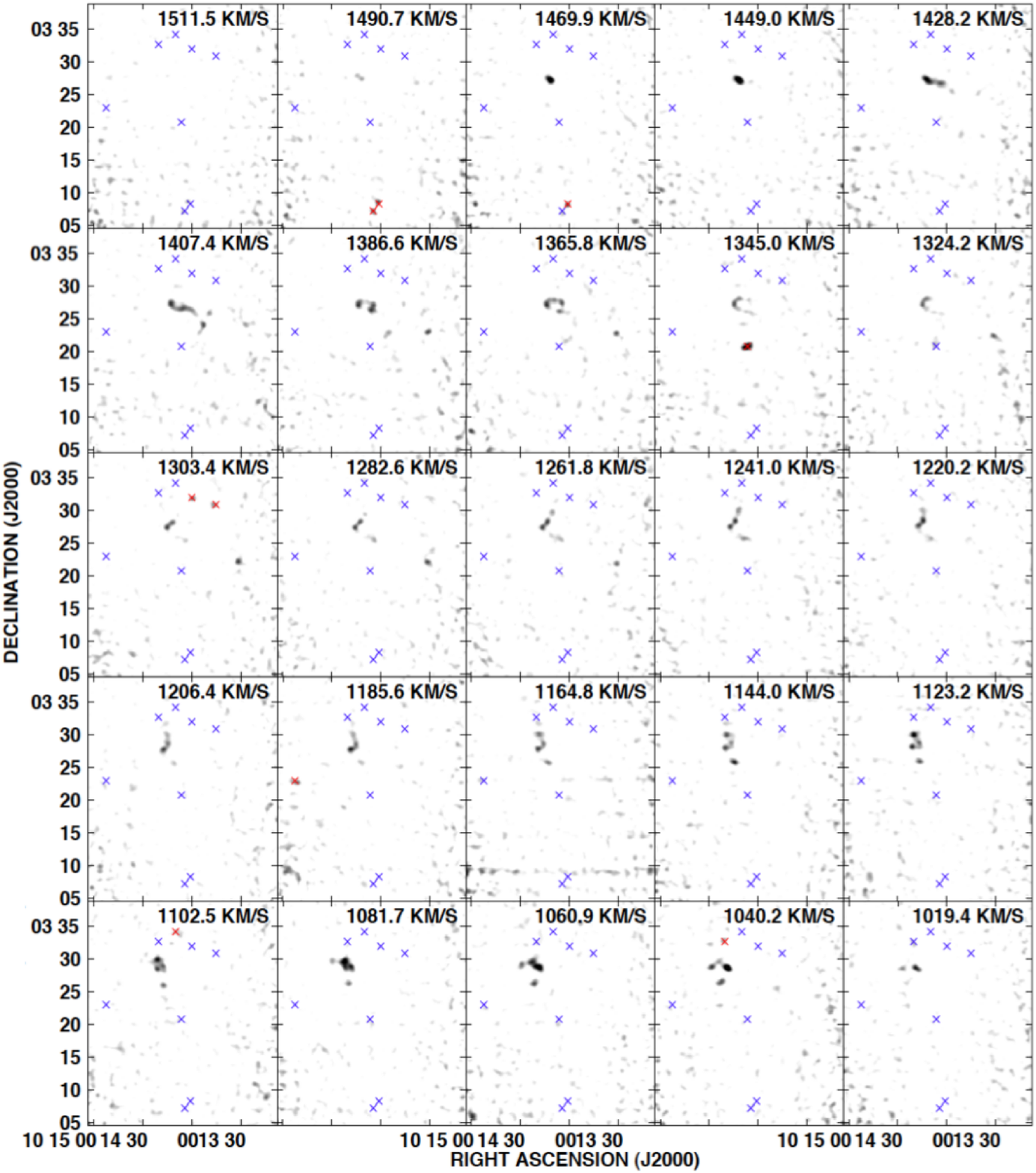}
  \caption{Channel maps of the mosaicked GMRT datacube at 45$\arcsec$ resolution.  For conciseness, only every third velocity channel is shown.  The locations of the detected low-mass objects are marked with blue ``X''s and turn red at the heliocentric velocity of the source (see Table~\ref{dyn}).  The grey scale runs from 3.6 to 24 mJy beam$^{-1}$ and the heliocentric recessional velocity of each channel is given in the top right corner of each panel.  The dark regions, representing bright emission, near $\alpha =$ 10\textsuperscript{h}14\textsuperscript{m}15\textsuperscript{s}, $\delta = +03^{\circ}28\arcmin00\arcsec$ are associated with NGC 3169.  \label{chan}}
\vspace{20mm}
\end{center}
\end{figure*}

\begin{figure*}
\begin{center}
  \includegraphics[width=84mm]{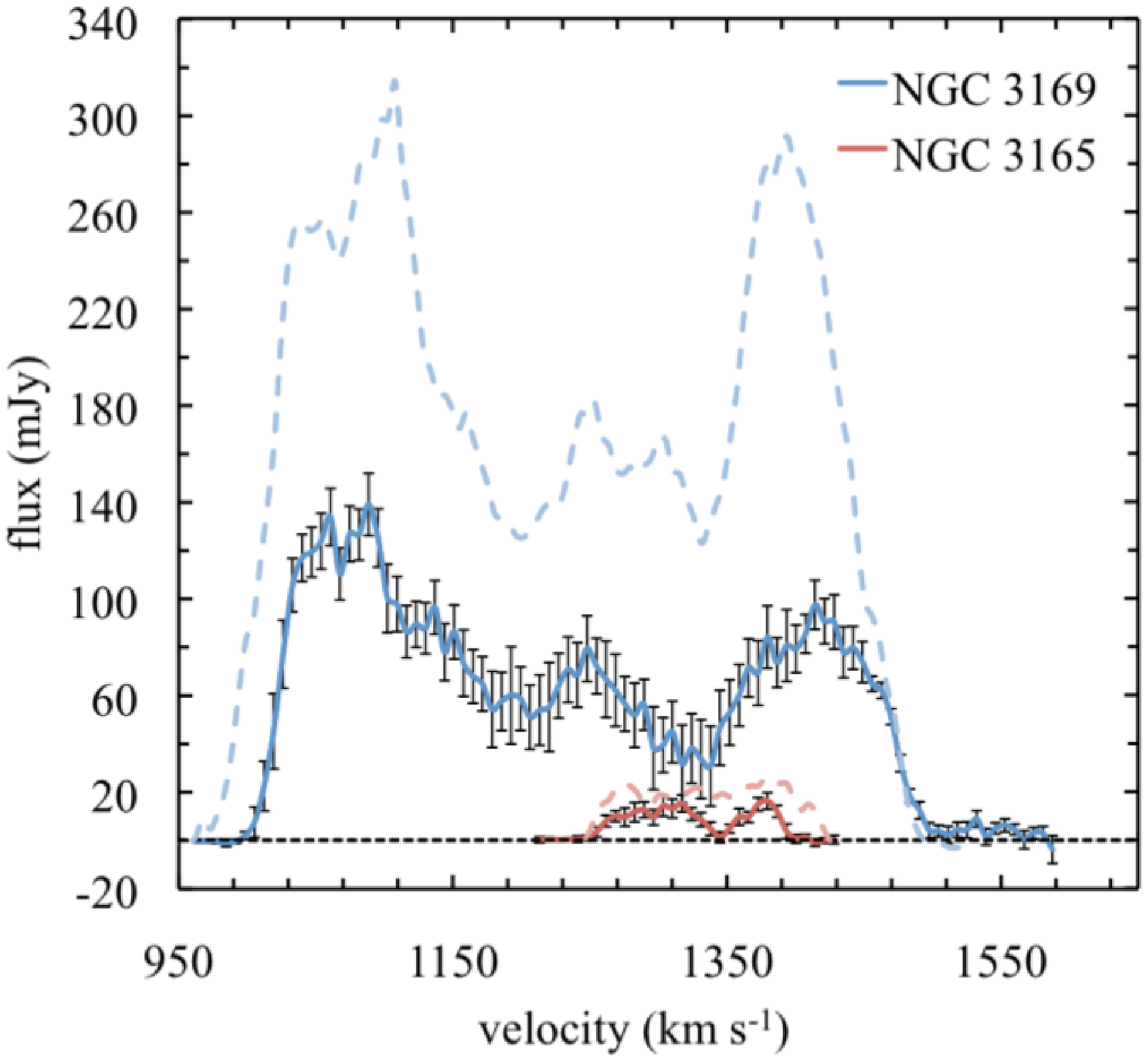}
    \includegraphics[width=84mm]{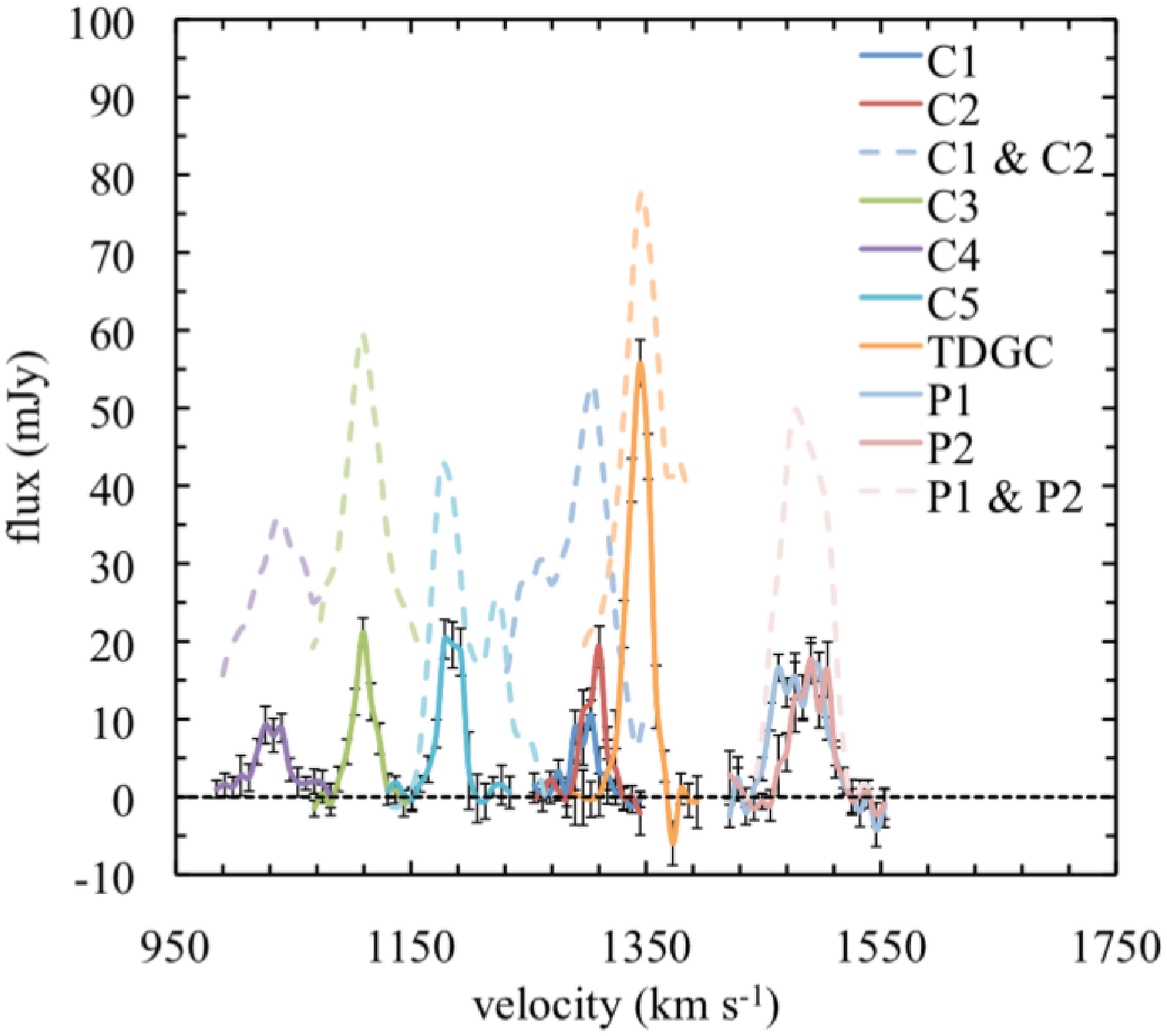}
  \caption{Global profiles derived from the ALFALFA (dashed lines) and GMRT (solid lines) data.  For clarity, the error bars for the ALFALFA data have been omitted.   \textit{Left:} NGC 3165 and NGC 3169.  \textit{Right:} Gas-rich, low-mass detections.  Some of the individual detections in the GMRT observations are spatially and spectrally unresolved in the ALFALFA data and have combined global profiles. \label{GP}}
\end{center}
\end{figure*}

We find ten {H~\sc{i}} sources in the data that meet our detection criteria. Their positions in the cube are illustrated in the channel maps of Fig.~\ref{chan}.  Their locations on the sky relative to the ALFALFA data and the GMRT data are shown in Figs.~\ref{point} and \ref{HI} respectively.  For clarity in Fig.~\ref{HI}, regions in each channel of the datacube where $\sigma>2.8$ mJy beam$^{-1}$ or where the signal-to-noise ratio S/N$<3$ were blanked before the moments shown were computed.  Fig.~\ref{GP} shows the global profiles of the {H~\sc{i}} detections measured from the 45$\arcsec$ resolution GMRT maps.  The measured {H~\sc{i}} properties of each detection are given in Tables~\ref{mass} and \ref{dyn}.  Despite good sensitivity in this region, we do not detect the arc to the NE of NGC 3169 at $\alpha =$ 10\textsuperscript{h}15\textsuperscript{m}16\textsuperscript{s}, $\delta = 3^{\circ}43\arcmin00\arcsec$ nor the enclosed AGC objects that are identified in the ALFALFA data (Fig.~\ref{ALFA}).  Given the integrated flux of AGC 204288, AGC 208329, AGC 208538 and AGC 208539 (Table~\ref{ALF_detect}), we expect their emission to fall below our GMRT detection threshold in that region if the {H~\sc{i}} emission is smoothly distributed over $\sim$3--10 GMRT beams.  Due to the large resolution difference between the ALFALFA and GMRT observations, these sources could still be unresolved in the ALFALFA maps as observed.  The same conclusions can be made for the other non-detections NW of NGC 3169: AGC 208533 and AGC 208534.  The ALFALFA detections AGC 5503, ACG 5539, AGC 204302, AGC 5551 and AGC 208392 are not in the GMRT mosaic.

NGC 3165 and NGC 3169 are gas-rich objects as shown in Fig.~\ref{HI}.  While the {H~\sc{i}} in NGC 3165 appears to be quite unperturbed, the optical photometry of \citet{j2000} shows significant asymmetries near the central region of the galaxy.  The role of NGC 3165 in the interaction is therefore unclear, although our data suggest that its contribution is minor and that this galaxy may have entered the group only recently.  There does appears to be some emission associated with NGC 3166 and the assumed starting location of an extended tidal feature, but it does not meet our detection criteria since it spans only two channels rather than the requisite three.  Nevertheless, the coincidence of this source with the optical position of NGC 3166 and the base of an {H~\sc{i}} tail strongly suggests that the emission is real (see left panel in Fig.~\ref{tail_tdg}).  A tidal bridge between the two largest spirals and a tidal tail containing $M_{H_I} \sim 1 \times 10^8$ $M_{\odot}$ extending below NGC 3166 are visible in Figs.~\ref{HI} and \ref{tail_tdg} at $\alpha =$ 10\textsuperscript{h}13\textsuperscript{m}55\textsuperscript{s}, $\delta = +03^{\circ}26\arcmin00\arcsec$ and $\alpha =$ 10\textsuperscript{h}13\textsuperscript{m}55\textsuperscript{s}, $\delta = +03^{\circ}23\arcmin00\arcsec$ respectively.  The complex morphology of the emission associated with NGC 3166 and NGC 3169 constrains the geometry and timing of their interaction but requires elaborate hydrodynamic simulations to determine the exact details (e.g.~\citealt{t1972}; \citealt{b1992}; \citealt{n2003}; \citealt{l2008}), which will be addressed in a later paper.  

\begin{figure}
  \includegraphics[width=84mm]{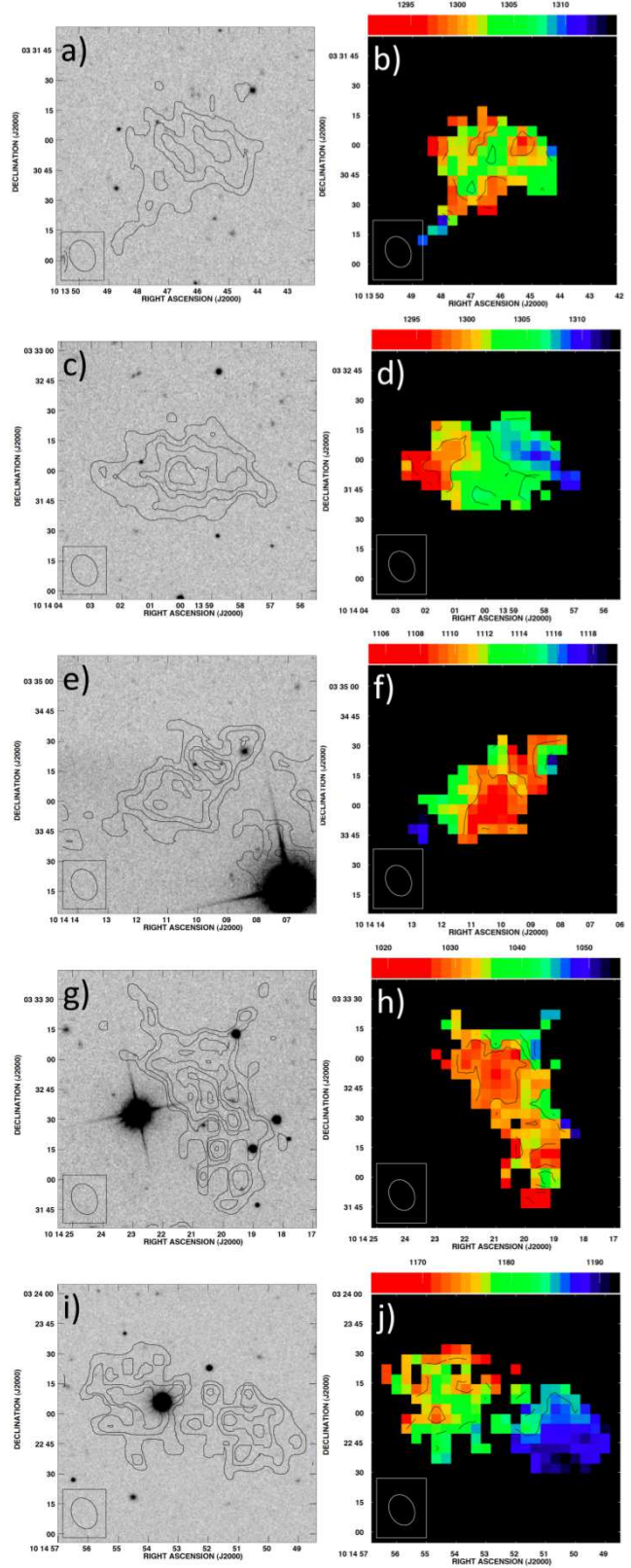}
  \caption{\textit{Left:} GMRT total intensity maps of sources C1 to C5 at 15$\arcsec$ resolution superimposed on SDSS  \textit{r}-band images.  Contours are at $N_{H_I} = (2.5, 3.75, 5, 6.25, 7.5) \times 10^{19}$ atoms cm$^{-2}$.  \textit{Right:} GMRT intensity-weighted velocity maps at 15$\arcsec$ resolution.  Velocity contours are at increments of 5 km s$^{-1}$.  a \& b) C1; c \& d) C2; e \& f) C3; g \& h) C4; i \& j) C5. \label{mom_detect}}
\end{figure}

\begin{figure}
  \includegraphics[width=84mm]{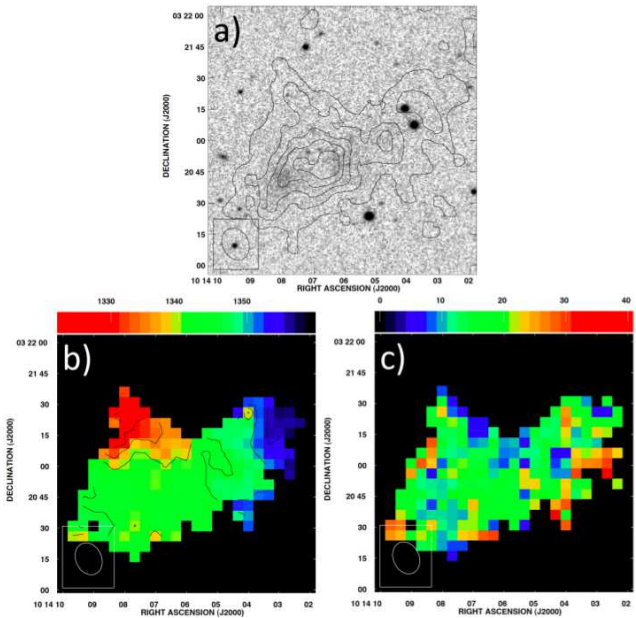}
  \caption{a) GMRT total intensity map of the TDG candidate at 15$\arcsec$ resolution superimposed on a SDSS  \textit{r}-band image.  Contours are at $N_{H_I} = (2.5, 5, 7.5, 10, 12.5) \times 10^{19}$ atoms cm$^{-2}$, which -- for clarity -- is twice the increment in comparison to the contours in Figs.~\ref{mom_detect} and \ref{mom_per}.  b) GMRT intensity-weighted velocity map at 15$\arcsec$ resolution.  Velocity contours are at (1325, 1330, 1335, 1340, 1345, 1350, 1355) km s$^{-1}$.  c) GMRT second velocity moment map at 15$\arcsec$ resolution. \label{mom_tdg}}
\end{figure}

\begin{figure}
  \includegraphics[width=84mm]{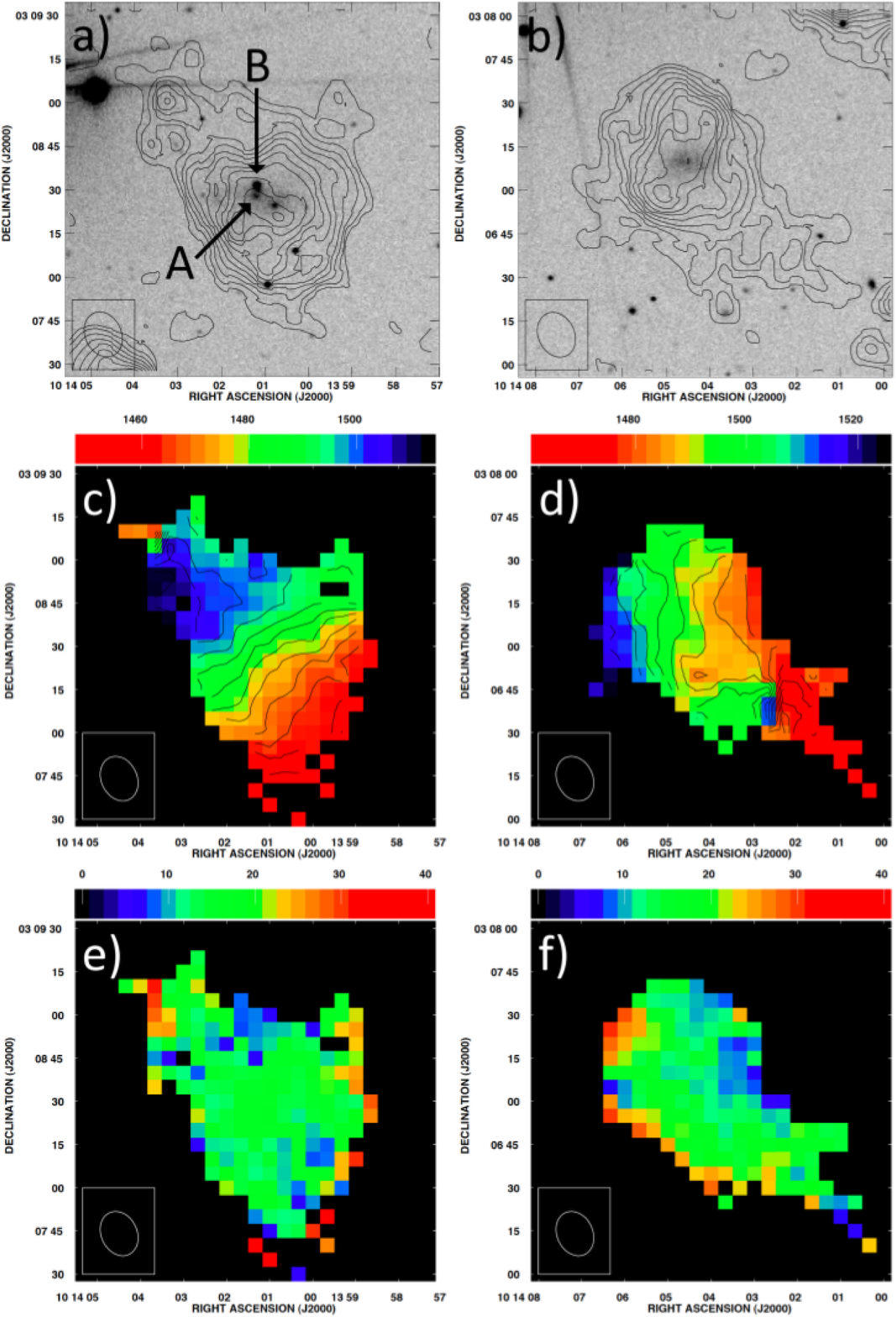}
  \caption{\textit{Top:} GMRT total intensity maps at 15$\arcsec$ resolution superimposed on a SDSS \textit{r}-band image.  Contours are at $N_{H_I} = (2.5, 3.75, 5, 6.25, 7.5, 8.75, 10, 11.25, 12.5, 13.75, 15) \times 10^{19}$ atoms cm$^{-2}$.  \textit{Centre:} GMRT intensity-weighted velocity maps at 15$\arcsec$ resolution.  Velocity contours are at increments of 5 km s$^{-1}$.  \textit{Bottom:} GMRT second velocity moment maps at 15$\arcsec$ resolution.  \textit{Left:} P1.  \textit{Right:} P2.  The labels A and B in panel a) denote the compact SDSS sources J101401.19+030827.7 and J101401.18+030831.6 respectively. \label{mom_per}}
\end{figure}

In addition to the gas-rich NGC galaxies, the GMRT observations reveal the structure of eight low-mass objects -- of which five were previously identified in the ALFALFA observations -- that are the primary focus of this paper.  Sources C1 to C5 and the TDG candidate, AGC 208457, are all found within the gas distribution in the group core (see Figs.~\ref{point} and \ref{HI}); whereas, P1 and P2 (AGC 208443 and AGC 208444) are situated in a separate, but spatially and spectrally close, region.  These two peripheral sources appear slightly outside of the velocity range of the larger group members, predominantly NGC 3169, while the other six detections have velocities within $\sim$1500 km s$^{-1}$ of NGC 3169.  Due to the large difference in angular resolution between the ALFALFA and GMRT observations, some distinct GMRT detections have not been identified in the ALFALFA data and do not have AGC numbers (see Table~\ref{mass}).  Accordingly, we refer to the low-mass {H~\sc{i}} sources by their designation in col.~1 of Table~\ref{mass}.  Total intensity and intensity-weighted velocity maps derived from the 15$\arcsec$ resolution datacubes are shown in Figs.~\ref{mom_detect}, \ref{mom_tdg} and \ref{mom_per} for sources C1 to C5, the TDG candidate, P1 and P2 respectively.  Since the signal-to-noise ratio of sources C1 to C5 is quite low and these five detections are unlikely to be in dynamical equilibrium, second velocity moment maps are included for the TDG candidate, P1 and P2 only (Figs.~\ref{mom_tdg} and \ref{mom_per}).

The TDG candidate is spatially located at the tip of a tidal tail that emerges from the vicinity of NGC 3166.  The velocity range of this feature also coincides with its assumed connection to the tail as there is a smooth velocity transition of the gas from the tidal tail into the TDG candidate.  Further evidence in favour of this object's tidal origin is discussed in $\S$ 6.

By assuming that {H~\sc{i}} is optically thin the {H~\sc{i}} mass, $M_{H_I}$, for each GMRT detection is calculated using:
\begin{equation}
M_{H_I} = 2.356 \times 10^5 d^2 S_{21} \hspace{3mm} M_{\odot}
\end{equation}
where $d$ is the distance to the source in Mpc and $S_{21}$ is the integrated flux density of the object in Jy km s$^{-1}$ \citep{g1988}, which we take as the integral of the global profiles in Fig.~\ref{GP} over velocity.  The uncertainty of $S_{21}$ was determined by propagating the uncertainties from the global profiles and adding, in quadrature, a 10\% flux calibration error.  For consistency -- and to ensure proper comparison for all GMRT detections -- we also calculated {H~\sc{i}} masses for the corresponding features of the ALFALFA cube using the same technique as adopted for the GMRT data.  The ALFALFA global profiles are shown in Fig.~\ref{GP} for comparison with the measurements from the GMRT data.  On account of its 4$\arcmin$ resolution, the ALFALFA observations are sensitive to large scale {H~\sc{i}} features while smaller structures remain unresolved.  Accordingly, some of the individual detections in the GMRT observations have only combined {H~\sc{i}} mass estimates from the ALFALFA data.  Both {H~\sc{i}} mass estimates are given in Table~\ref{mass}.  

\begin{table*}
\centering
 \begin{minipage}{170mm}
\caption{{H~\sc{i}} Masses from the GMRT and ALFALFA data}
\label{mass}
\begin{tabular}{@{}l c c c @{}c @{} r@{}l @{} r@{}l @{} r@{}l @{}} 
\hline
Source 	&AGC	&GMRT Coordinates		&Optical Coordinates	&$\sigma$	&\multicolumn{2}{c}{$S_{21}$}		&\multicolumn{2}{c}{$M_{H_I}$}		&\multicolumn{2}{c}{$M'_{H_I}$}\\
		&		&(J2000) 				&(J2000)		&(mJy beam$^{-1}$)	&\multicolumn{2}{l}{(Jy km s$^{-1}$)}	&\multicolumn{2}{l}{($10^7 M_{\odot}$)}	&\multicolumn{2}{c}{($10^8 M_{\odot}$)}\\
(1)		&(2)		&(3) 					& (4) 				& (5)		&\multicolumn{2}{c}{(6)}			&\multicolumn{2}{c}{(7)}				&\multicolumn{2}{c}{(8)}\\	
\hline
NGC 3165 		&5512	&10 13 31.0, +03 22 30	&10 13 31.3, +03 22 30	& 2.55	&\hspace{4mm}1.5 &\hspace{1mm}$\pm$ 0.2 	&\hspace{3mm}18 &\hspace{1mm}$\pm$ 3 			& \hspace{8.5mm}3.8 &\hspace{1mm}$\pm$ 0.8\\
NGC 3169 		&5525	&10 14 15.0, +03 28 00	&10 14 14.8,	+03 27 59	& 1.70	& 35 &\hspace{1mm}$\pm$ 4 		& 420 &\hspace{1mm}$\pm$ 50 		& 110 &\hspace{1mm}$\pm$ 10\\
C1 		&		&10 13 45.3, +03 30 53	&					& 1.95		& 0.26 &\hspace{1mm}$\pm$ 0.06 	& 3.1 &\hspace{1mm}$\pm$ 0.8		&\multicolumn{2}{c}{combined with C2}\\
C2 		&		&10 14 00.0, +03 31 58	&					& 1.80		& 0.40 &\hspace{1mm}$\pm$ 0.08 	& 5 &\hspace{1mm}$\pm$ 1			& 2.4 &\hspace{1mm}$\pm$ 0.3\\ 
C3 		&208535 &10 14 10.0, +03 34 10	&					& 1.80		& 0.4 &\hspace{1mm}$\pm$ 0.1 	& 5 &\hspace{1mm}$\pm$ 1 			& 2.4 &\hspace{1mm}$\pm$ 0.5\\
C4 		&		&10 14 20.4, +03 32 40	&					& 1.70		& 0.29 &\hspace{1mm}$\pm$ 0.06 	& 3.5 &\hspace{1mm}$\pm$ 0.8 		& 1.8 &\hspace{1mm}$\pm$ 0.4\\
C5 		&208537 &10 14 52.5, +03 23 00 	&					& 2.15		& 0.53 &\hspace{1mm}$\pm$ 0.09 	& 6 &\hspace{1mm}$\pm$ 1 			& 1.4 &\hspace{1mm}$\pm$ 0.3\\	
TDGC	&208457 &10 14 06.5, +03 20 47	&10 14 08.0,	+03 20 41	& 1.90	& 1.9 &\hspace{1mm}$\pm$ 0.3 	& 23 &\hspace{1mm}$\pm$ 3 			& 3.8 &\hspace{1mm}$\pm$ 0.6\\
P1		&208443 &10 14 01.0, +03 08 18 	&10 14 01.0,	+03 08 27	& 2.40	& 0.9 &\hspace{1mm}$\pm$ 0.1 	&10 &\hspace{1mm}$\pm$ 2 			&\multicolumn{2}{c}{combined with P2}\\
P2		&208444 &10 14 04.5, +03 07 13 	&10 14 04.6, +03 07 10	& 2.45	& 0.6 &\hspace{1mm}$\pm$ 0.1 	&7 &\hspace{1mm}$\pm$ 1			& 2.9 &\hspace{1mm}$\pm$ 0.4\\
\hline 
\multicolumn{11}{l}{Parameters measured from the GMRT are unprimed; whereas, those measured from the ALFALFA data are primed.}\\
\multicolumn{11}{l}{\footnotesize Col. (1) detection name}\\
\multicolumn{11}{l}{\footnotesize Col. (2) corresponding AGC number}\\
\multicolumn{11}{l}{\footnotesize Col. (3) right ascension and declination of peak {H~\sc{i}} flux density}\\
\multicolumn{11}{l}{\footnotesize Col. (4) centroid of the most probable optical counterpart, where applicable}\\
\multicolumn{11}{l}{\footnotesize Col. (5) RMS noise in region of detection in GMRT maps}\\
\multicolumn{11}{l}{\footnotesize Col. (6) integrated {H~\sc{i}} flux density computed from the global profiles in Fig.~\ref{GP}.  The uncertainty was determined by propagating the}\\
\multicolumn{11}{l}{\footnotesize \hspace {1cm} error from the global profile and adding a 10\% calibration error}\\
\multicolumn{11}{l}{\footnotesize Col. (7) {H~\sc{i}} mass calculated from GMRT observations using Eq. 1}\\
\multicolumn{11}{l}{\footnotesize Col. (8) {H~\sc{i}} mass from ALFALFA observations calculated using the same technique that was used to determine {H~\sc{i}} mass from the}\\
\multicolumn{11}{l}{\footnotesize \hspace {1cm} GMRT observations.  Unresolvable objects are combined into one measurement}\\
\end{tabular}
\end{minipage} 
\end{table*}

Overall, the GMRT maps recover $<40\%$ of the ALFALFA flux density (Fig.~\ref{GP} and Table~\ref{mass}), which suggests that a significant fraction of the {H~\sc{i}} in the group is diffuse and smoothly distributed throughout the system on arcminute (7 kpc) scales.  Of the ten GMRT detections, five can be correlated with AGC objects.  The ALFALFA values that we computed for NGC 3169 agree, within error, to the values presented for AGC 5525 in Table~\ref{ALFA}.  The spatial and spectral locations of C3, C5, the TDG candidate, P1 and P2 correspond to AGC 208535, AGC 208537, AGC 208457, AGC 208443 and AGC 208444 respectively.  For these five low-mass objects, our calculation method -- which constrains the spatial size of these objects using the GMRT maps, unlike the $7\arcmin \times 7 \arcmin$ integration region used by ALFALFA to determine $S'_{21}$ \citep{h2011} -- recovers $\sim$60\% of the total flux presented in Table~\ref{ALFA}.  This lower flux in the GMRT detections compared to their ALFALFA counterparts is consistent with the non-detection of other ALFALFA sources in the group core at higher resolution.

Under the assumption that the low-mass GMRT detections are in dynamical equilibrium, an estimate of their total dynamical masses in the region where {H~\sc{i}} is detected is given by:
\begin{equation}
M_{dyn} = 3.39 \times 10^4 a_{H_I} d \left( \frac{W_{20}}{2} \right)^2 \hspace{3mm} M_{\odot}
\end{equation}
where $a_{H_I}$ is the diameter of the object in arcminutes, $d$ is the distance to the source in Mpc and $W_{20}$ is the width, at 20\% of the peak flux density, of its global profile in km s$^{-1}$ \citep{g1988}.  No correction for inclination is made to $W_{20}$ in Eq.~2.  The major axis diameter of each detection was measured from the total intensity maps in Figs.~\ref{mom_detect} - \ref{mom_per} and then adjusted for beam smearing effects -- by deconvolving the length with a gaussian with a FWHM equivalent to the HPBW of the beam -- to yield an estimate of $a_{H_I}$.  Either $W_{20}$ or $W_{50}$, the width at 50\% of the global profile peak, can be used to compute dynamical masses \citep{r1994}.  Due to the fact that the detections of interest have relatively narrow linewidths and because we wish to conservatively estimate their dynamical masses so as to place an upper limit on their dark matter fractions (see $\S$ 5), $W_{20}$ was adopted for our calculations.  For example, using $W_{50}$ for the detections with the lowest dynamical masses would imply a 50\% lower mass than that computed with $W_{20}$, which would decrease the dynamical to baryonic mass fraction by the same amount.  On account of the discrepancy between the usage of $W_{20}$ or $W_{50}$, the uncertainties for the dynamical masses were taken as half the difference between the masses computed using each value.  The total dynamical masses and other related quantities are given in Table~\ref{dyn}.  Our dynamical masses are generally consistent with those estimated from the maximum velocities at each end of the axis used to measure $a_{H_I}$ in the velocity moment map; however, since very few of these detections show signs of rotation, we adopted the more conservative method of using $W_{20}$ from the integrated spectra.

Note that the assumption of dynamical equilibrium is likely inappropriate for C1-C4.  In the interest of completeness and in keeping with the unbiased approach of our study, we derive the same basic properties for all the GMRT detections.  We then discuss the likely origin of the detections in $\S$6.

\begin{table}
 \centering
 \begin{minipage}{75mm}
\caption{Dynamical Masses from the GMRT Data}
\label{dyn}
\begin{tabular}{@{} l  c c c r@{}l} 
\hline
Source 		& W$_{20}$ 			&$a_{H_I}$		&$cz_{\odot}$		&\multicolumn{2}{c}{$M_{dyn}$} \\
			& (km s$^{-1}$)		&(arcmin)		&(km s$^{-1}$)	&\multicolumn{2}{c}{($10^8 M_{\odot}$)} \\
(1)			& (2) 				& (3)			&(4)				&\multicolumn{2}{c}{(5)}	\\
\hline
C1 			& 36 $\pm$ 7 		& 1.1 $\pm$ 0.1	& 1302 $\pm$ 3	& 2.6 &\hspace{1mm}$\pm$ 0.8 \\
C2 			& 33 $\pm$ 7 		& 1.4 $\pm$ 0.1	& 1307 $\pm$ 3	& 3.0 &\hspace{1mm}$\pm$ 0.9 \\
C3 			& 34 $\pm$ 7 		& 1.2 $\pm$ 0.1	& 1110 $\pm$ 3	& 3 &\hspace{1mm}$\pm$ 1 	\\
C4			& 54 $\pm$ 7 		& 1.6 $\pm$ 0.1	& 1028 $\pm$ 3	& 9 &\hspace{1mm}$\pm$ 3 	\\
C5 			& 33 $\pm$ 7 		& 1.9 $\pm$ 0.1	& 1182 $\pm$ 3	& 4 &\hspace{1mm}$\pm$ 1 	\\
TDGC		& 33 $\pm$ 7 		& 2.1 $\pm$ 0.1	& 1343 $\pm$ 3	& 4 &\hspace{1mm}$\pm$ 1 	\\
P1		 	& 66 $\pm$ 7 		& 1.7 $\pm$ 0.1	& 1482 $\pm$ 3	& 14 &\hspace{1mm}$\pm$ 3 \\
P2			& 51 $\pm$ 7 		& 1.7 $\pm$ 0.1	& 1488 $\pm$ 3	& 8 &\hspace{1mm}$\pm$ 2 	\\
\hline
\multicolumn{6}{l}{\footnotesize Col. (1) detection name}\\
\multicolumn{6}{l}{\footnotesize Col. (2) global profile width measured at 20\% of the}\\
\multicolumn{6}{l}{\footnotesize \hspace{1cm} peak flux density}\\
\multicolumn{6}{l}{\footnotesize Col. (3) beam corrected major axis diameter of object}\\
\multicolumn{6}{l}{\footnotesize \hspace{1cm} in 15$\arcsec$ resolution maps}\\
\multicolumn{6}{l}{\footnotesize Col. (4) heliocentric velocity of the profile midpoint at}\\
\multicolumn{6}{l}{\footnotesize \hspace{1cm} 20\% of the peak flux density}\\
\multicolumn{6}{l}{\footnotesize Col. (5) total dynamical mass calculated using Eq. 2.}\\
\multicolumn{6}{l}{\footnotesize \hspace {1cm} The uncertainty is half the difference between }\\
\multicolumn{6}{l}{\footnotesize \hspace {1cm} $M_{dyn}$ computed with $W_{20}$ and $W_{50}$.}\\
\end{tabular}
\end{minipage} 
\end{table}


\section[]{Ancillary Data}

To further understand the nature of the gas-rich, low-mass objects identified in the {H~\sc{i}} observations, archival optical and UV datasets, from the Sloan Digital Sky Survey (SDSS DR7; \citealt{a2009}) and the Galaxy Evolution Explorer (GALEX GR6; \citealt{m2005}), were respectively utilized to estimate stellar masses and SFRs of these systems.  After reviewing archival SDSS images in the various bands, the more sensitive \textit{r}-band images were used to estimate stellar masses.  In Figs.~\ref{mom_detect} - \ref{mom_per}, the 15$\arcsec$ resolution GMRT total intensity maps of each detection are plotted on top of SDSS \textit{r}-band images of the same region.  For clarity, the total intensity contours of the TDG candidate (Fig.~\ref{mom_tdg}a) are plotted at twice the increment of the contours in Figs.~\ref{mom_detect} and \ref{mom_per}.  

\begin{figure*}
\begin{center}
  \includegraphics[width=84mm]{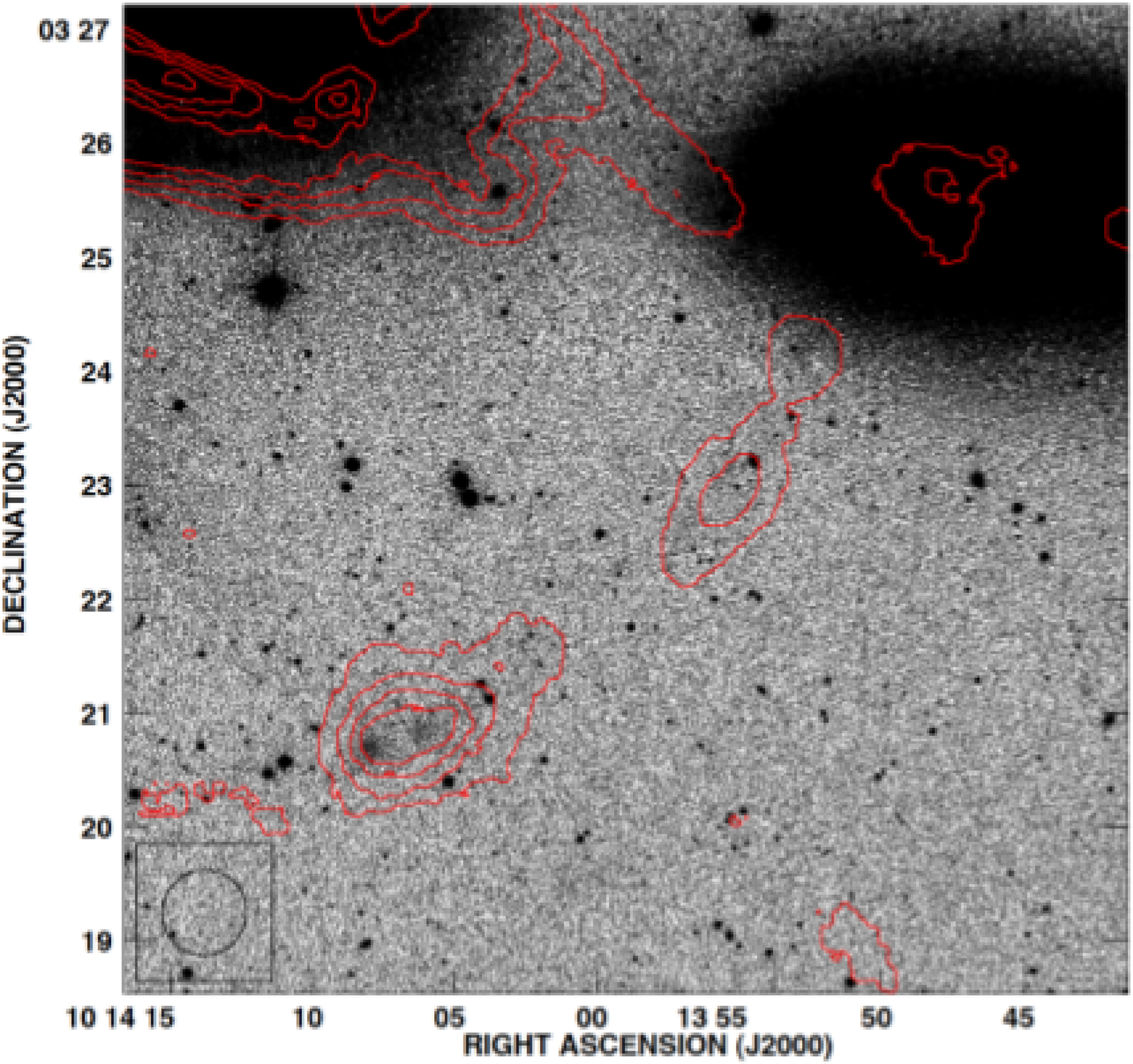}
    \includegraphics[width=84mm]{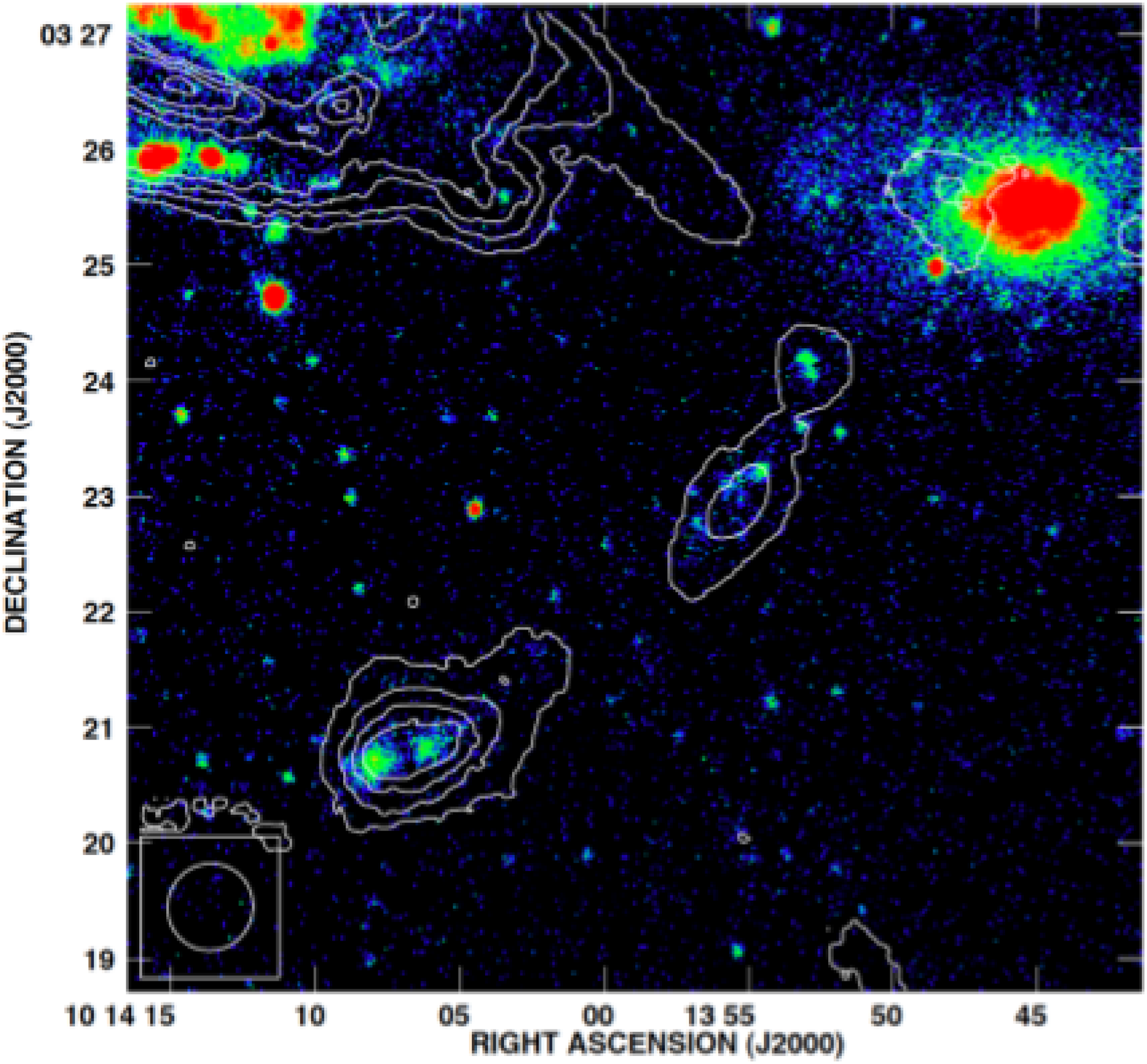}
  \caption{GMRT total intensity map of the tidal tail -- at 45$\arcsec$ angular resolution -- extending below NGC 3166 and towards the TDG candidate.  Contours are at $N_{H_I} = (0.8, 2, 3, 4) \times 10^{20}$ atoms cm$^{-2}$.  \textit{Left:} {H~\sc{i}} contours superimposed on a SDSS \textit{r}-band image.  The large dark objects (representing bright features) are NGC 3166 and NGC 3169.  There appears to be some very faint optical emission in the vicinity of the tidal tail and the TDG candidate.  \textit{Right:} {H~\sc{i}} contours superimposed on a GALEX NUV image.  The presumed UV tidal tail is closely followed by the {H~\sc{i}} contours that culminate on a region of active star formation with a relatively high {H~\sc{i}} density. \label{tail_tdg}}.  
\end{center}
\end{figure*}

The TDG candidate, P1 and P2 all appear to coincide with faint, extended optical features that are well below the 5$\sigma$ significance threshold of the SDSS catalogue \citep{a2009}; whereas, sources C1 to C5 have no distinct optical counterparts (see also Table~\ref{ALF_detect}).  Fig.~\ref{mom_per}a also includes two SDSS detections in the vicinity of the optical feature that is possibly associated with the {H~\sc{i}} detection P1.  J101401.19+030827.7 and J101401.18+030831.6 (labelled A and B) have reported sizes of 1.54$\arcsec$ $\times$ 1.28$\arcsec$ and 6.28$\arcsec$ $\times$ 3.79$\arcsec$ \citep{a2007} respectively and appear to contribute minimally to the diffuse emission.  We used the underlying low surface brightness emission (likely shredded by the SDSS pipeline) around these SDSS detections in our calculations.  The left panel of Fig.~\ref{tail_tdg} shows a portion of the 45$\arcsec$ resolution {H~\sc{i}} data superimposed on a SDSS \textit{r}-band image.  There appears to be a very faint linear feature in the SDSS image that underlies the tidal tail identified in the GMRT data.  There is also diffuse optical emission south of NGC 3169 at $\alpha =$ 10\textsuperscript{h}14\textsuperscript{m}15\textsuperscript{s}, $\delta = +03^{\circ}23\arcmin00\arcsec$ in the SDSS image.  This region lies well within the {H~\sc{i}} envelope detected by ALFALFA but does not appear to be associated with either an ALFALFA or a GMRT detection (see Figs.~\ref{ALFA} and \ref{point}).

Two extended optical features of unknown redshift are symmetrically located around the {H~\sc{i}} peak of the TDG candidate and appear at roughly the SDSS \textit{r}-band RMS noise limit, $\sigma_s$ (Fig.~\ref{mom_tdg}).  We presume that both features are associated with the TDG candidate.  In the fields of sources P1 and P2 (Fig.~\ref{mom_per}), the assumed optical counterparts are at approximately twice the noise level.  According to \citet{y2000}, the 5$\sigma_s$ detection limit is 23.1 magnitudes for point sources in 1$\arcsec$ seeing using the SDSS $r$ filter that is equivalent to an \textit{r}-band flux of $F_\lambda = 1.544 \times 10^{-17}$ W m$^{-2}$ $\mu$m$^{-1}$ \citep{f1996}.  Therefore, the RMS noise is $\sigma_s = 3.088 \times 10^{-18}$ W m$^{-2}$ $\mu$m$^{-1}$.  We compute a total \textit{r}-band flux for each detection by multiplying its estimated optical surface area with either the RMS noise level, $\sigma_s$, in the case for the TDG candidate or twice the limit, $2\sigma_s$, for P1 and P2.  For the five detections without optical counterparts, a circular source with a diameter of 10$\arcsec$ -- which was assumed to have a flux per pixel value of $\sigma_s$ -- was chosen.  This method provided a conservative estimate of the luminosity in the \textit{r}-band, $L_r$.  A characteristic value of the stellar mass to stellar light ratio, $M/L_r$ $\sim$ 1.6, was used to estimate the stellar mass from $L_r$ \citep{b2003}.  

Table~\ref{final} shows stellar mass estimates obtained using the method described above as well as gas mass estimates derived from the {H~\sc{i}} data.  Considering that the vast majority of the hydrogen in low-mass galaxies is atomic (e.g.~\citealt{l2005}), $M_{H_I}$ is a fair indicator of total gas mass, $M_{gas}$, which is given by:
\begin{equation}
M_{gas} = 1.33 M_{H_I}
\end{equation}
where the factor of 1.33 accounts for the mass from helium and other elements.  The stellar masses were added to the gas masses of each detection to estimate the total baryonic mass, $M_{baryon}$, where a 50\% uncertainty was assumed for the stellar mass estimates.  

Following the same approach as outlined for the SDSS measurements, we examined archival GALEX observations in the vicinity of the low-mass {H~\sc{i}} detections.  Again, the TDG candidate, P1, P2 and the tidal tail appear to have relatively bright UV counterparts at both wavelengths while sources C1 to C5 have no discernible UV emission.  The right panel of Fig.~\ref{tail_tdg} shows the 45$\arcsec$ resolution {H~\sc{i}} data of the tidal tail region superimposed on a GALEX NUV image.  The {H~\sc{i}} in the tail closely follows a linear UV feature.  The coincidence of the {H~\sc{i}} in the tidal tail with both the optical and UV features that are found in the same region strongly suggests that all these features are at the same redshift. 

For each of the three low-mass objects with discernible UV counterparts, a counts per second (CPS) value for each pixel was determined by averaging five independent measurements of the systems in the GALEX images.  These values were then converted into fluxes, F$_{FUV}$ and F$_{NUV}$, using:
\begin{equation}
\mbox{F$_{FUV}$ [erg s$^{-1}$ cm$^{-2}$ \AA$^{-1}$]} = 1.40 \times 10^{-15} \times \mbox{ CPS}
\end{equation}
\begin{equation}
\mbox{F$_{NUV}$ [erg s$^{-1}$ cm$^{-2}$ \AA$^{-1}$]} = 2.06 \times 10^{-16} \times \mbox{ CPS}
\end{equation}
\citep{m2007} and the corresponding luminosities were converted to SFRs using the following relation:
\begin{equation}
\mbox{SFR$_{UV}$ ($M_{\odot}$ yr$^{-1}$)} = \frac{L_{UV} \mbox{(erg s$^{-1}$ Hz$^{-1}$)}}{\eta_{UV}}
\end{equation}
where $\eta_{UV}$ depends on the initial mass function of the stellar population (\citealt{r2002}; \citealt{t2007}).  In the ``continuous star formation'' approximation, $\eta_{UV} \sim \eta^0 _{UV} = 1.4\times 10^{28}$ erg yr s$^{-1}$ Hz$^{-1}$ M$_{\odot} ^{-1}$ \citep{k1998}.  As there was only a slight variation between the computed SFRs in the FUV and the NUV, an average of the results, SFR$_{\overline{UV}}$, was computed to represent an overall SFR, which is presented in Table~\ref{final}.  

\begin{table*}
 \centering
 \begin{minipage}{120mm}
\caption{Properties of the Detected Low-mass Objects in NGC 3166/9}
\label{final}
\begin{tabular}{@{}l r@{}l c r@{}l r@{}l r@{}l c} 
\hline 
Source 	&\multicolumn{2}{c}{$M_{gas}$}	&$M_{stellar}$		&\multicolumn{2}{c}{$M_{baryon}$}		&\multicolumn{2}{c}{$M_{dyn}/M_{gas}$}		
	&\multicolumn{2}{c}{$M_{dyn}/M_{baryon}$}	&SFR$_{\overline{UV}}$\\
		&\multicolumn{2}{c}{($10^7 M_{\odot}$)}				&($10^7 M_{\odot}$)					&\multicolumn{2}{c}{($10^7 M_{\odot}$)	}	
	&\multicolumn{2}{c}{}	&\multicolumn{2}{c}{}							&($M_{\odot}$ yr$^{-1}$)\\
(1)		&\multicolumn{2}{c}{(2)} 			&(3)					&\multicolumn{2}{c}{(4)}				&\multicolumn{2}{c}{(5)}						
	&\multicolumn{2}{c}{(6)}						&(7)	\\
\hline 	
C1 	&\hspace{1mm} 4 &\hspace{1mm}$\pm$ 1 	& $<0.2$		&\hspace{1mm} 4 &\hspace{1mm}$\pm$ 1	&\hspace{3mm} 6 &\hspace{1mm}$\pm$ 4				
& 6 &\hspace{1mm}$\pm$ 4					& $<0.001$	\\
C2 		& 6 &\hspace{1mm}$\pm$ 1 		& $<0.2$			& 7 &\hspace{1mm}$\pm$ 1				& 5 &\hspace{1mm}$\pm$ 2				
& 5 &\hspace{1mm}$\pm$ 2					& $<0.001$	\\
C3 		& 7 &\hspace{1mm}$\pm$ 2 		& $<0.2$			& 7 &\hspace{1mm}$\pm$ 2				& 4 &\hspace{1mm}$\pm$ 2				
& 4 &\hspace{1mm}$\pm$ 2					& $<0.001$	\\
C4 		& 5 &\hspace{1mm}$\pm$ 1 		& $<0.2$			& 5 &\hspace{1mm}$\pm$ 1				& 20 &\hspace{1mm}$\pm$ 10			
& 20 &\hspace{1mm}$\pm$ 10				& $<0.001$	\\
C5 		& 9 &\hspace{1mm}$\pm$ 1 		& $<0.2$			& 9 &\hspace{1mm}$\pm$ 1		& 5 &\hspace{1mm}$\pm$ 2				
&\hspace{5mm} 5 &\hspace{1mm}$\pm$ 2		& $<0.001$	\\ 
TDGC	& 31 &\hspace{1mm}$\pm$ 4 		& $\sim1$			& 32 &\hspace{1mm}$\pm$ 4				& 1.5 &\hspace{1mm}$\pm$ 0.6			
& 1.4 &\hspace{1mm}$\pm$ 0.4				& $\sim$0.01	\\
P1		& 14 &\hspace{1mm}$\pm$ 2 		& $\sim1$			& 15 &\hspace{1mm}$\pm$ 2				& 10 &\hspace{1mm}$\pm$ 4			
& 9 &\hspace{1mm}$\pm$ 4					& $\sim$0.02	 \\
P2		& 9 &\hspace{1mm}$\pm$ 2		& $\sim1$			& 10 &\hspace{1mm}$\pm$ 2				& 10 &\hspace{1mm}$\pm$ 4			
& 8 &\hspace{1mm}$\pm$ 4					& $\sim$0.005 \\
\hline 
\multicolumn{11}{l}{\footnotesize Col. (1) detection name}\\
\multicolumn{11}{l}{\footnotesize Col. (2) gas mass estimated using the GMRT {H~\sc{i}} mass and Eq. 3}\\
\multicolumn{11}{l}{\footnotesize Col. (3) stellar mass estimated from archival SDSS \textit{r}-band data}\\
\multicolumn{11}{l}{\footnotesize Col. (4) baryonic mass}\\
\multicolumn{11}{l}{\footnotesize Col. (5) dynamical to gas mass ratio}\\
\multicolumn{11}{l}{\footnotesize Col. (6) dynamical to baryonic mass ratio}\\
\multicolumn{11}{l}{\footnotesize Col. (7) star formation rate estimated from archival GALEX data}\\
\end{tabular}
\end{minipage} 
\end{table*}

\subsection{Notes on Individual Objects}

Using a uniform method for the analysis, the {H~\sc{i}} properties of the eight low-mass detections were measured in $\S$ 4 and their stellar masses, baryon fractions and SFRs were estimated above.  As mentioned previously, all dynamical masses have been computed under the strong assumption that each detection is self-gravitating in dynamical equilibrium, which is probably inaccurate for some objects (see $\S$ 6).  Here, we discuss the observed properties that will be used to constrain the origins of the individual detections in $\S$ 6.
\vspace{2mm}

\noindent C1 is the weakest GMRT detection as it has a peak {H~\sc{i}} flux density at 4.7$\sigma$.  It is located in close proximity to C2, which appears in the same velocity channels.  The {H~\sc{i}} in and around C1 may form a gaseous filament, which includes C2, that extends away from NGC 3169.  Based on the assumption that C1 is in dynamical equilibrium, this detection might be dark matter dominated with $M_{dyn}/M_{gas} = 6 \pm 4$; as discussed in $\S$ 6, however, this interpretation is unlikely to be true.
\vspace{2mm}

\noindent C2 appears to have a smooth velocity gradient in Fig.~\ref{mom_detect}d that could reflect the motion of the gas within the surrounding filament.  If self-gravitating, this detection appears to be dark matter dominated with $M_{dyn}/M_{gas} = 5 \pm 2$.  There seems to be an association between C1 and C2; however, further observations at higher sensitivity are required to better define their relationship.    
\vspace{2mm}

\noindent C3 corresponds to AGC 208535 in the ALFALFA data.  The GMRT data recovers $\sim$20\% of flux measured from the ALFALFA data implying that there is a significant amount of {H~\sc{i}} distributed on arcminute scales in this region.  Assuming that C3 is in dynamical equilibrium with $M_{dyn}/M_{gas} = 4 \pm 2$, then this object has the least amount of dark matter of the five low-mass core region GMRT detections.  The velocity distribution in Fig.~\ref{mom_detect}f suggests largely turbulent motions but there is a hint that the outer regions (bottom left and top right) of the gas in this detection are being pulled towards other high density areas. 
\vspace{2mm}

\noindent C4 is one of the weakest detections as it has a peak {H~\sc{i}} flux density just below 5$\sigma$.  It appears to have very turbulent gas motions and shows no sign of rotation (Fig.~\ref{mom_detect}h).  Its proximity to NGC 3169 suggests that some of its gas may have been stripped by the larger galaxy.  If self-gravitating, C4 has $M_{dyn}/M_{gas} = 20 \pm 10$ and would be the most dark matter dominated gas-rich, low-mass object of the entire group.  As discussed in $\S$ 6, however, this assumption is not likely valid for C4.
\vspace{2mm}

\noindent C5 is located near the outer edge of the GMRT {H~\sc{i}} map, as seen in Fig.~\ref{HI}, and corresponds to AGC 208537 in the ALFALFA data.  The GMRT data recover less than half the flux measured from the ALFALFA data implying that there is a significant amount of {H~\sc{i}} distributed on arcminute scales in this region.  The smooth velocity gradient and lobed structure seen in Fig.~\ref{mom_detect}i,j suggests that the gas content in C5 has yet to be significantly affected by the group environment.  Accordingly, it has the highest estimated gas mass, $M_{gas} = 9 \pm 1 \times 10^7$ $M_{\odot}$, of the five low-mass core region detections.  C5 appears to be dark matter dominated with $M_{dyn}/M_{gas} = 5 \pm 2$. 
\vspace{2mm}

\noindent The TDG candidate has a relatively high peak {H~\sc{i}} flux density and with $M_{dyn}/M_{gas} = 1.5 \pm 0.6$, it contains little to no dark matter.  The GMRT observations of this object recover $\sim$$60\%$ of the {H~\sc{i}} detected by ALFALFA; nevertheless, when the missing {H~\sc{i}} is included, dynamical to baryonic mass ratio remains consistent with unity.  Fig.~\ref{mom_tdg}b,c illustrates that there is no clear velocity gradient across the object and the second velocity moment has a roughly constant value of 20 km s$^{-1}$ that is similar in amplitude to W20 (Table~\ref{dyn}). This finding suggests that the TDG is not rotationally supported.  The TDG candidate corresponds to AGC 208457 in the ALFALFA data and emerges at the tip of a tidal tail that is discernible in the optical and in the UV.  From ancillary optical and UV data, there appears to be traces of both old and young stellar populations associated with this {H~\sc{i}} detection.   
\vspace{2mm}

\noindent P1 and P2 have a fair amount of gas and each is associated with faint optical emission, indicative of stellar populations, in Fig.~\ref{mom_per}a,c.  Both are equally dark matter dominated with $M_{dyn}/M_{gas} = 10 \pm 4$.  There appears to be a modest amount of on-going star formation in these gas-rich systems and their smooth velocity gradients in Fig.~\ref{mom_per}b,d suggest that these objects are rotating.  The second velocity moments in Fig.~\ref{mom_per}e,f are roughly constant across each source, with a median value $\sim$15 km s$^{-1}$.  This value further suggests that these objects are rotation-dominated.  The proximity and common velocity range of P1 and P2 suggests that the pair are interacting (a part of P2 is seen in the bottom left corner of Fig.~\ref{mom_per}a and the top right corner of Fig.~\ref{mom_per}c shows a part of P1).  These two detections correspond to AGC 208443 and AGC 208444 in the ALFALFA data.  Overall, the GMRT data recovers $\sim$60\% of the combined flux measured from the ALFALFA data for these two sources, which implies that a portion of the {H~\sc{i}}  in this region is distributed on arcminute scales.  Although these two galaxies are not in the NGC 3166/9 core region (see Fig.~\ref{point}), their close proximity to the group -- both spatially and in velocity -- suggests a physical association.


\section[]{Discussion and Conclusions}

We have presented ALFALFA and follow-up GMRT {H~\sc{i}} observations of the gas-rich interacting group NGC 3166/9.  The sensitive ALFALFA data provide a complete census of the {H~\sc{i}}-bearing systems in the group while the high-resolution GMRT data elucidate the origin of the eight systems detected with both instruments. The combination of ALFALFA and GMRT observations therefore enables one of the first unbiased studies of pre-exisiting dwarfs, tidal knots and tidal dwarf galaxies in nearby groups.  The ALFALFA maps reveal an extended {H~\sc{i}} envelope around the group core.  We identified ten sources within this core region in our higher resolution GMRT observations, of which two sources correspond to NGC 3165 and NGC 3169.  The other eight detections are gas-rich, low-mass objects.  Additionally, a tidal bridge between the two largest spirals, NGC 3166 and 3169, as well as a tidal tail extending below NGC 3166 were identified in the GMRT datacube.  The tidal tail below NGC 3166 was shown to coincide with similar features in both SDSS and GALEX maps of the region.  For each of the low-mass objects, we have computed gas masses, dynamical masses, stellar masses and SFRs.  Below, we discuss the potential origin of each system given these properties.

Most of our detections do not appear to have significant optical or UV counterparts, which suggests that the majority of the baryons in these systems are in the form of atomic gas.  P1 and P2 show signs of having faint stellar disks that are currently experiencing active star formation.  As well, there is a very faint optical feature and a more pronounced UV detection in Fig.~\ref{tail_tdg} that coincide with the tidal tail and the TDG candidate suggesting that stars are being extracted from the larger spirals while recently stripped gas is undergoing star formation.

\begin{figure*}
\begin{center}
  \includegraphics[width=84mm]{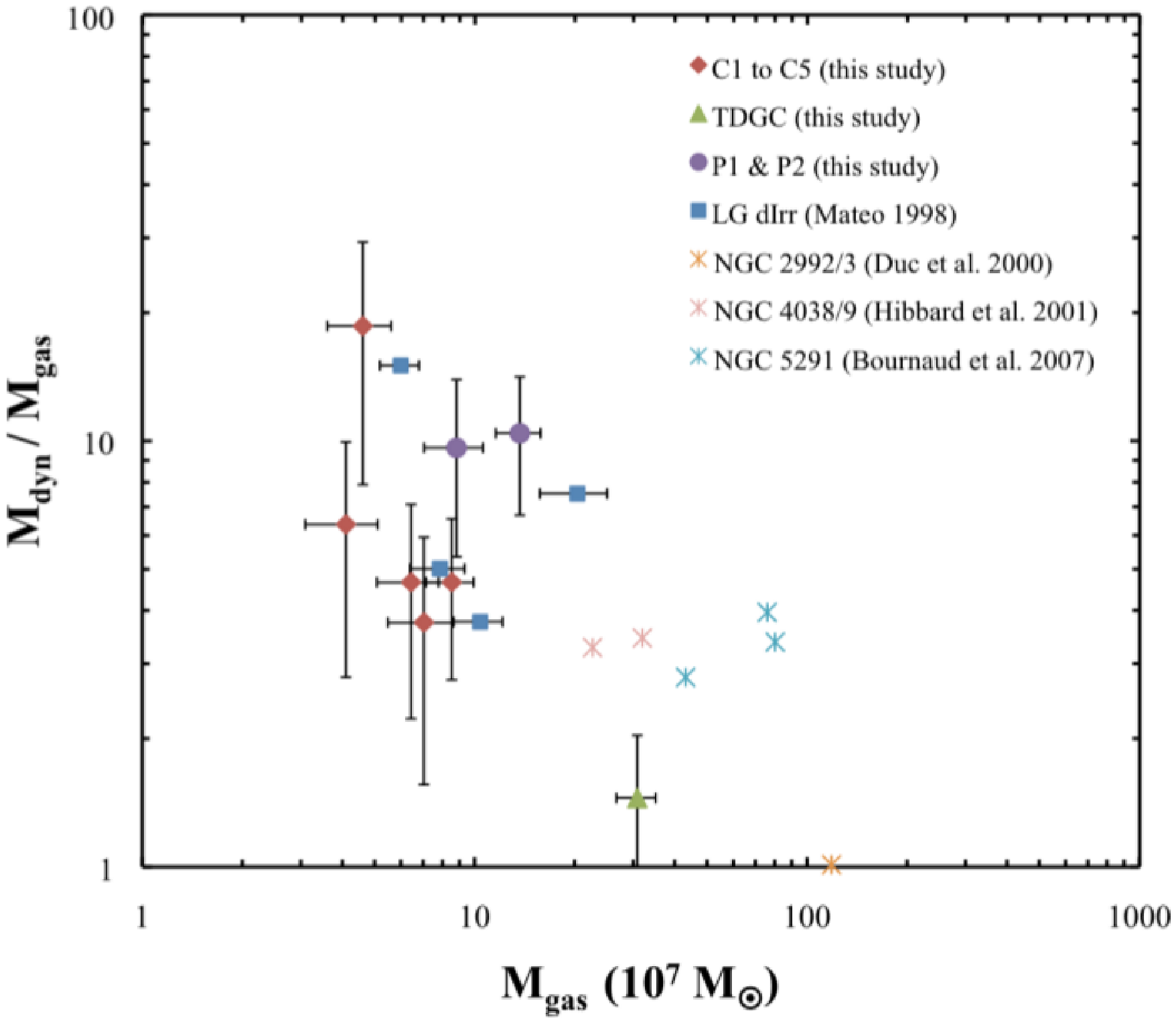}
    \includegraphics[width=84mm]{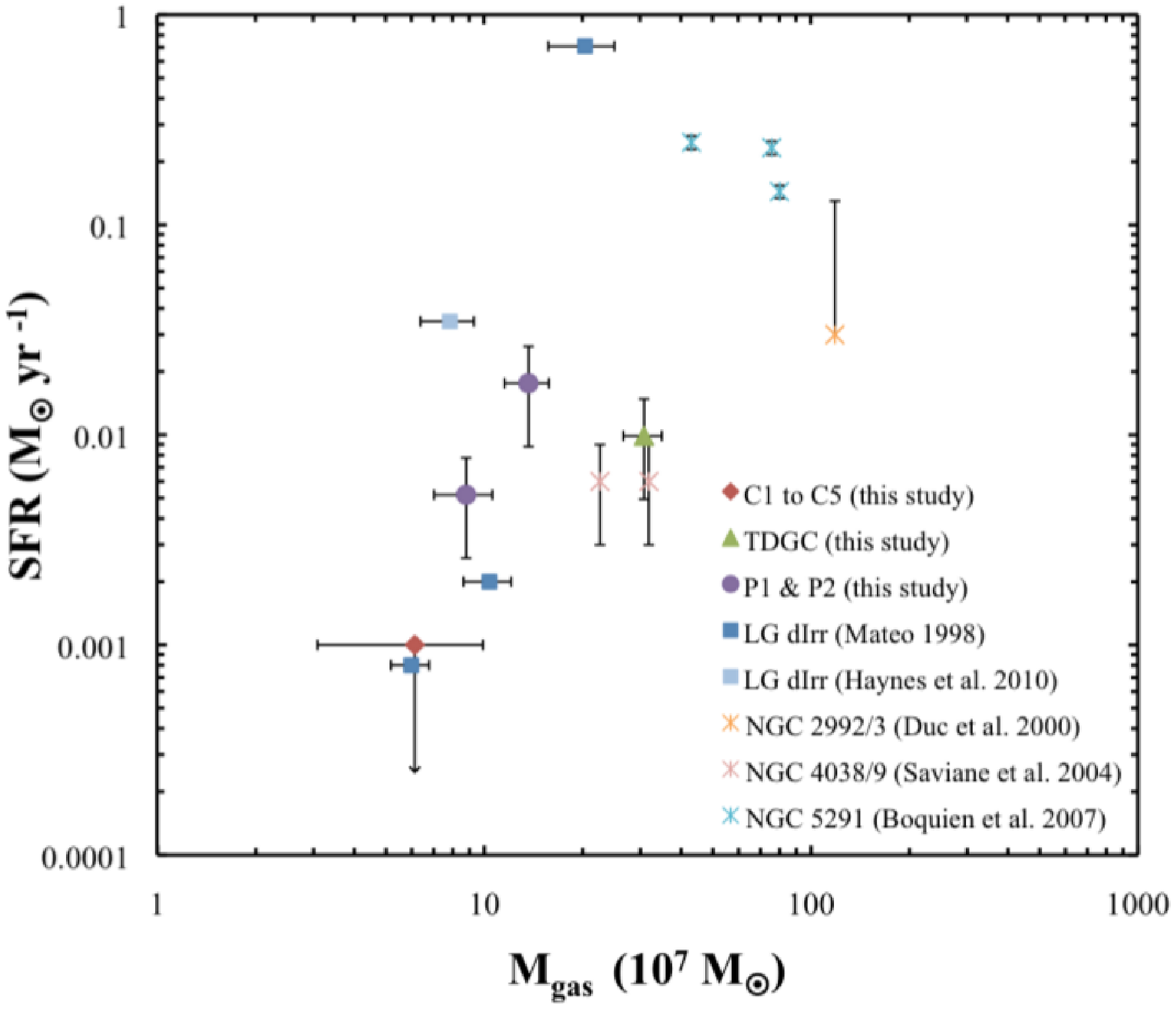}
  \caption{\textit{Left:} Mass ratios for the low-mass detections in this study (diamonds, circles and triangle), dIrrs in the LG (squares; as classifed by \citet{m1998}) and other TDG candidates (asterisks) from the literature.  It is likely that C1 to C4, included for comparison purposes, are not self-gravitating and their dynamical mass estimates have no physical meaning.  $M_{gas}$ for each of the LG dIrrs was computed from the data compiled by \citet{m1998} and was compared to the reported total masses typically measured from the width of the 21-cm line.  The appropriate masses for the TDG candidates A245N, S78, MDL92, NGC5291N, NGC5291S and NGC5291SW were computed from VLA data by the authors indicated on the plot.  \textit{Right:} Current SFRs for various gas-rich objects.  The SFRs computed for sources C1 to C5 are upper limits and represented by a single point (diamond) plotted as the mean with the error bar for $M_{gas}$ representing the standard deviation of the individual means.  SFRs for the LG dwarfs were estimated by \citet{m1998} using extinction-corrected H$\alpha$ fluxes.  Where H$\alpha$ information was unavailable, the SFR was estimated based on GALEX FUV integrated photometry by \citet{h2010}.  SFRs for the TDG candidates were estimated using various methods by the authors indicated on the plot.  In both graphs, error bars are only plotted for the values that included error estimates. \label{LG}}
\end{center}
\end{figure*}

Fig.~\ref{LG} compares the mass ratios and SFRs of the low-mass objects in this study, the dwarf irregular galaxies (dIrrs) in the Local Group (LG) with $M_{H_I} > 2 \times 10^7$ $M_{\odot}$ (the GMRT detection limit at the map centre) and a small sample of TDG candidates from the literature whose properties were also determined using high-resolution {H~\sc{i}} observations (see below).  Eq. 3 was used to compute $M_{gas}$ for the LG galaxies, which were then compared to the reported total masses in the compiled data from \citet{m1998}.  There are several claims of TDG candidates in current literature (e.g.~\citealt{b2001}; \citealt{d2007}; \citealt{c2009}; \citealt{h2009}); nevertheless, there are very few unambiguous identifications.  The candidates chosen for comparison in this study are the particularly well-known -- and well-observed -- TDG candidates that have {H~\sc{i}} and dynamical masses computed from interferometric {H~\sc{i}} observations: A245N in NGC 2992/3 from \citet{d2000}, S78 and MDL92 in NGC 4038/9 -- originally detected by \citet{s1978} and \citet{m1992a} -- observed in {H~\sc{i}} by \citet{h2001} and NGC5291N, NGC5291S and NGC5291SW around NGC 5291, which were recently observed by \citet{b2007} and \citet{boq2007}.

There appears to be a rough grouping distinguishing dIrrs from TDG candidates in Fig.~\ref{LG}.  Since the GMRT observations are sensitive to all gas-rich objects in the mass range shown -- unless they were to appear at the edge of the mosaic -- the segregation by gas mass is likely not a selection effect.  Sources C1 to C5, P1 and P2 all have gas masses and dynamical to gas mass ratios similar to several dIrrs found in the LG; whereas, the TDG candidates have comparatively higher gas contents and lower mass ratios.  The right panel of Fig.~\ref{LG} shows that LG dIrrs generally have minimal amounts of current star formation with the exception of IC 10, which exhibits complex kinematics, a disturbed outer {H~\sc{i}} velocity field and an unusually high SFR that suggests an ongoing interaction \citep{m1998}.  Due to the similarity of (and uncertainty in) their SFRs (Table~\ref{final}), sources C1 to C5 are represented by a single point.  Most of our detections have values corresponding to SFRs of LG dIrrs while the TDG candidates appear to have slightly higher SFRs.  

The value of $M_{dyn}/M_{gas}$ for C5 is similar to that of typical dIrrs.  The low stellar to gas mass ratio and smooth lobed structure of C5 indicates that it has not been affected by the interaction between the larger NGC galaxies; whereas, C4 appears to have been significantly affected by the gravity of NGC 3169, which has resulted in a fairly low gas mass and very turbulent gas motions for the object.  Judging by their spatial and spectral distribution as well as their proximity to NCG 3169, there is a possibility that C1, C2 and C3 are located in a gaseous tail of tidal origin; however, according to \citet{b2006}, with $M_{gas}$ $\sim 7 \times 10^{7}$ $M_{\odot}$ these objects are unlikely to be long-lived and would be classified as transient tidal knots, which will eventually fall back onto their progenitor.  In this context, it is likely that sources C1 to C4 are not self-gravitating and therefore that their dynamical mass estimates in Table 4 have no physical meaning.  The apparent signs of current and previous star formation as well as their potential involvement with NGC 3166/9 warrants more detailed study of probable dIrrs P1 and P2 that should lead to a better understanding of galaxy pre-processing around group environments (see \citealt{l2009}).

The TDG candidate in NGC 3166/9 exhibits the typical dynamical signatures of other likely second-generation dwarfs (see Fig.~\ref{LG}).  The channel maps in Fig.~\ref{chan} show that the velocity of this object defines its clear association with the tidal tail that extends below NGC 3166.  At the same time, it is clearly distinct from the tail in {H~\sc{i}} and optical brightness as well as in morphology.  Its high gas mass content, $M_{gas} = 3.1 \pm 0.4 \times 10^8$ $M_{\odot}$, and a low dynamical to gas mass ratio, $M_{dyn}/M_{gas}$ = $1.5 \pm 0.6$, is in agreement with the properties of predicted and observed tidally formed dwarf galaxies (i.e. \citealt{d2000}; \citealt{b2007}).  The TDG candidate appears to have a very faint optical component in Fig.~\ref{mom_tdg}a and the associated UV emission in Fig.~\ref{tail_tdg} suggests that there is a population of newly formed stars.  

Recent simulations and observations suggest that at least some TDGs should show signs of rotation \citep{d2007}.  There does appear to be some form of coherent motion within the TDG candidate in NGC 3166/9 in Fig.~\ref{mom_tdg}b; however, a velocity gradient that is consistent with rotation is not confirmed and the amplitude of its velocity width ($W_{20} \approx 35$ km s$^{-1}$) is similar to the velocity dispersion across the object implied by the second velocity moment map in Fig.~\ref{mom_tdg}c.  These results suggest that the TDG is not rotationally supported, contrary to other {H~\sc{i}}-rich TDGs so far studied in the literature and the predictions from simulations \citep{bo2008}.  The lack of clear rotation does reflect the findings for other TDG candidates (such as that in NGC 4038/9; \citealt{h2001}); whereas, the few TDG candidates that do show convincing signs of rotation are significantly more massive than the one reported in this paper (see \citealt{b2007}).  Only in this respect do the properties of the TDG candidate differ from some model predictions \citep{d2011}.  Nevertheless, its location at the tip of a tidal tail, its gas and stellar content, its SFR and most importantly its baryon fraction are all consistent with both simulations and the properties of other TDG candidates, which strongly suggest a tidal origin for this object.

Additional observations -- such as optical spectroscopy to estimate metallicity and confirm a rotational velocity gradient as well as deep optical imaging to better probe stellar populations -- are needed verify the hypothesis that the TDG candidate discussed here is indeed a second-generation dwarf (e.g.~\citealt{detal2011}).  Numerical simulations constrained to reproduce the global {H~\sc{i}} properties of the NGC 3166/9 group could also provide insight into the origin of the system.  We are exploring all of these potential avenues for NGC 3166/9 as well as other gas-rich groups probed by ALFALFA as part of a larger campaign to determine both the prevalence and properties of tidally formed structures in nearby group environments.


\vspace{8mm}
We thank the staff of the GMRT for facilitating our interferometric observations and the many ALFALFA team members who contributed to producing the data used here.  Thank-you to the reviewer, P.-A. Duc, for his numerous suggestions to improve the clarity of this paper.  K. S. acknowledges funding from the National Sciences and Engineering Research Council of Canada.  The ALFALFA team at Cornell is supported by grants from the U.S. National Science Foundation NSF/AST-0607007 and AST-1107390 and by a grant from the Brinson Foundation.

The GMRT is run by the National Centre for Radio Astrophysics of the Tata Institute of Fundamental Research.  The Arecibo Observatory is operated by SRI International under a cooperative agreement with the National Science Foundation (AST-1100968) and in alliance with Ana G. M\'endez-Universidad Metropolitana and the Universities SpaceResearch Association.  This research made use of Montage, funded by the National Aeronautics and Space Administration's Earth Science Technology Office, Computation Technologies Project, under Cooperative Agreement Number NCC5-626 between NASA and the California Institute of Technology. Montage is maintained by the NASA/IPAC Infrared Science Archive.  Some of the data presented in this paper were obtained from the Multimission Archive at the Space Telescope Science Institute (MAST). STScI is operated by the Association of Universities for Research in Astronomy, Inc., under NASA contract NAS5-26555. Support for MAST for non-HST data is provided by the NASA Office of Space Science via grant NNX09AF08G and by other grants and contracts.  This research has also made use of the NASA/IPAC Extragalactic Database (NED) which is operated by the Jet Propulsion Laboratory, California Institute of Technology, under contract with the National Aeronautics and Space Administration.

\end{document}